\documentclass[useAMS,usenatbib,referee]{biom}
\def\bSig\mathbf{\Sigma}

\usepackage{amsmath, amsfonts, amssymb, array}

\title[``Robust-squared'' Imputation]{``Robust-squared'' Imputation Models Using BART}

\author{Yaoyuan V. Tan$^{1,*}$\email{vincetan@umich.edu},
Carol A.C. Flannagan$^{2}$, and Michael R. Elliott$^{1}$ \\
$^{1}$Department of Biostatistics, University of Michigan, Ann Arbor, U.S.A \\
$^{2}$University of Michigan Transportation Research Institute, University of Michigan, Ann Arbor, U.S.A}

\begin{document}

\date{{\it Received } . {\it Revised } .\newline
{\it Accepted } .}

\pagerange{\pageref{firstpage}--\pageref{lastpage}} \pubyear{}

\volume{}
\artmonth{}
\doi{10.1111/j.1541-0420.2005.00454.x}

%  This label and the label ``lastpage'' are used by the \pagerange
%  command above to give the page range for the article

\label{firstpage}

%  pub the summary here

\begin{abstract}
Examples of ``doubly robust'' estimator for missing data include augmented inverse probability weighting (AIPWT) models \citep{rrz} and penalized splines of propensity prediction (PSPP) models \citep{zhang_little}. Doubly-robust estimators have the property that, if either the response propensity or the mean is modeled correctly, a consistent estimator of the population mean is obtained. However, doubly-robust estimators can perform poorly when modest misspecification is present in both models \citep{ks}. Here we consider extensions of the AIPWT and PSPP models that use Bayesian Additive Regression Trees \cite[BART;][]{chipman_bart} to provide highly robust propensity and mean model estimation.
We term these ``robust-squared'' in the sense that the propensity score, the means, or both can be estimated with minimal model misspecification, and applied to the doubly-robust estimator.
We consider their behavior via simulations where propensities and/or mean models are misspecified. We apply our proposed method to impute missing instantaneous velocity (delta-v) values from the 2014 National Automotive Sampling System Crashworthiness Data System dataset and missing Blood Alcohol Concentration values from the 2015 Fatality Analysis Reporting System dataset. We found that BART applied to PSPP and AIPWT, provides a more robust and efficient estimate compared to PSPP and AIPWT, with the BART-estimated propensity score combined with PSPP providing the most efficient estimator with close to nominal coverage.
\end{abstract}

%
%  Please place your key words in alphabetical order, separated
%  by semicolons, with the first letter of the first word capitalized,
%  and a period at the end of the list.
%


\maketitle

\noindent
{\bf{1. Introduction}}
\label{s:intro}

Missing data are common in many studies, surveys, and experiments. Data may be missing because of the subject's refusal to provide information or study drop-out, or by the design of the experiment or study. If the amount of missing data is large, or if the missing data differ from the observed data and would change our conclusions if we had observed it, failure to account for missing data during analysis leads to biased parameter estimation and misleading conclusions.
Missing data in surveys, including major US transportation safety-related surveys, is very common. The National Automotive Sampling System -- Crashworthiness Data System (NASS-CDS) is representative of all police-reported towaway crashes in the US. A key measure of crash severity is the ``instantaneous'' change in velocity, delta-v.  Because estimation of delta-v requires a careful crash investigation that is not always possible, it is commonly missing.  Similarly, the Fatality Analysis Reporting System (FARS) releases information annually from all fatal motor vehicle crashes that occur on US public roads.  Here, blood alcohol concentration (BAC) levels are often missing because subjects were not tested at the crash site.

To determine the best course of action to handle missing data, it is essential to recognize the mechanism of the missing data. In general, the missingness mechanism can be separated into three categories: missing completely at random (MCAR), where the data are missing by chance and are not related to observed or unobserved variables; missing at random (MAR), where the data are missing depending on some variables which are fully observed; and not missing at random (NMAR), where the data are missing depending on the variable that contains the missing value. In our examples, delta-v is often missing in vehicles that have either quite limited damage (so that the vehicle may be been driven off and not available for followup) or very severe damage (so that the algorithms used to estimate it do not have reliable inputs); while this might seem to imply NMAR, there are a number of observed measures such as towaway status, injury severity, and speed limit to make the MAR assumption more plausible.  Similarly, BAC measures are often missing in subjects that did not appear to be intoxicated; again factors such as gender, age, time of day, and crash severity can strengthen what, without other covariates, would seem to imply an NMAR mechanism.  Since MAR assumptions do not typically need to rely on unobservable parameters and can be reasonable given sufficient fully observed covariates,
 MAR is a common assumption that researchers adopt and shall be the focus of this paper.

A common way to handle missing data under MAR is to impute the missing values. There are many methods developed to handle missingness under MAR. Common methods include mean imputation, regression imputation, and hot deck \cite[][Chapter 4]{little_rubin}. Once the missing values are imputed, standard statistical techniques can be employed as though there were no missingness in the dataset. Valid inference accounts for the additional uncertainty due to imputation. This is commonly achieved using multiple imputation (MI) where $D$ imputed datasets are generated and the within and between variability of the estimator are calculated and combined to give the total uncertainty of the imputed estimator \cite[See][Chapter 5 ]{little_rubin}. Alternatively, bootstrap could be used to estimate the additional uncertainty created by the imputation method. Both methods rely on modeling assumptions that might not be correct or difficult to test.

\cite{rrz}, proposed a method, augmented inverse probability estimator (AIPWT), that separately modeled the response propensity and mean of the outcome as a function of observed data, and yields a consistent estimator if either model is specified correctly. Another robust method is the penalized splines of propensity prediction (PSPP) proposed by \cite{little_an} and later modified by \cite{zhang_little}. This method, while based on a Bayesian prediction framework different relative to the method proposed by Robins et al., has the same property that either a correct specification of response propensity or mean model produces consistent estimates, and thus are usually called doubly robust (DR) estimators. Extensions to multiply robust estimators that allow multiple models to be specified and yield consistent estimates as long as at least one is correct have been developed as well \citep{han}.

Unfortunately, DR estimators may not work that well in situations where both the propensity model and mean model are misspecified. \cite{ks} showed this using a simulation example where both the propensity and mean model were moderately misspecified. AIPWT and PSPP did worse in terms of bias compared to a method that only used a mean model for imputation. For real-life datasets, of course, true models are almost never known. Thus, modifying the AIPWT and PSPP so that these methods are robust to misspecification of both the propensity and mean model becomes important for AIPWT and PSPP to remain relevant outside theoretical and simulation settings.

Current literature modifying DR estimators so that they become robust to misspecification mainly focus on two observations. First, the propensity model can produce large weights and hence cause severely biased estimates in DR estimators when both propensity and mean models are incorrectly specified. Second, the propensity and mean model are misspecified because of the non-linear main and multiple-way interaction effects.

For the former observation, \cite{ks} proposed to replace the logistic regression with the robit regression \citep{liuc} where the robit regression replaces the logistic link with the Student-t distribution. On the other hand, \cite{cao} recognized that good performance of the propensity model in AIPWT relies on the summation of the multiplication of the propensity score and response being close to the sample size. Hence, they suggested estimating the logistic regression model with the restriction that the summation of the multiplication of the propensity score and response is approximately equal to the sample size. More recently, \cite{imai} proposed the covariate balancing propensity score (CBPS) where they focused on balancing the moments of the covariates between missing and non-missing groups instead of searching for a better parametric approach.

%For the later observation, researchers mainly propose to replace the method for propensity model estimation with one which could easily handle the non-linear main and multiple-way interaction effects without much setup by the researcher. These examples include using a generalized boosted model \citep{ridgeway}, classification and regression tree (CART), and random forest \citep{lee}.

In this paper, we capitalize on the fact that the PSPP model is already robust to the misspecification of the mean model, since it only requires that residuals of the misspecified mean model be a smooth function of the probability of non-response. Hence, a robust estimator of the response propensity will yield an estimator with especially strong robustness properties. Specifically, we estimate of the propensity model using Bayesian additive regression trees (BART) \citep{chipman_bart}. BART models the conditional mean of $\mathbf{Y}$ given $X$ as a sum of regression trees. Use of regression trees allows automatic incorporation of multi-way interactions; non-linear main effects and multi-way interactions can be incorporated through the summation of these trees.

The use of BART to replace the usual imputation model is not entirely new. \cite{xu} suggested using BART as the imputation model for situations where there is sequential missingness and it was possible to write the likelihood function in a sequential fashion. \cite{kapelner_miss} on the other hand suggested an approach to impute missing the regression trees if there are missingness in the covariates used to generate the regression trees. The novelty of our work is the combination of the AIPWT or PSPP model with BART to create a doubly-robust model where the degree of the misspecification for the estimation of the propensity model is greatly reduced: hence, our ``robust-squared' terminology. We also consider the use of BART in estimating the mean model as well.

We organize the rest of our manuscript as follows. In Section 2, we describe our missing data problem followed by a brief review of the AIPWT, PSPP, and BART. We present our proposed methods for extending AIPWT and PSPP followed by suggesting two imputation methods using BART directly in Section 3. In Section 4, we employ a simulation study to compare our proposed methods against AIPWT and PSPP. In Section 5, we compared various imputation methods on the estimation of the population mean of delta-v and unadjusted odds ratio of injury severity using the 2014 National Automotive Sampling System Crashworthiness Data System (NASS-CDS) dataset as well as estimation of the population mean of Blood Alcohol Concentration (BAC) and proportion of subjects with BAC more than .010 and .100 using the 2015 Fatality Analysis Reporting System (FARS) dataset. Section 6 concludes with a discussion and possible future work.

\noindent
{\bf{2. Review of existing Doubly Robust methods for MAR data}}

\noindent
{\bf{2.1 Description and Notation}}

Suppose we have a scalar variable $Y_k$, $k=1,\ldots,n$ and we are interested in estimation and inference of $E[\mathbf{Y}]=\mu$, the population mean. We will consider continuous $Y_k$, but extensions to binary and categorical $Y_k$ are certainly possible. Let $R_k=1$ denote that the $k^{\text{th}}$ element of $\mathbf{Y}$ is observed and $R_k=0$ denote that the $k^{\text{th}}$ element is missing. We restrict to situations where missingness of $Y_k$ depends on $p$ fully-observed covariates $\mathbf{X}_k=(X_{k1},\ldots,X_{kp})^T$.

\noindent
{\bf{2.2 Robins, Rotnitzky, Zhao (1994) augmented inverse probability estimator (AIPWT)}}.

To address the missing data problem described above, \cite{rrz} proposed a double robust estimator by solving a set of estimating equations. In brief, $\mu$ is estimated as
\begin{equation}
	\label{rrz}
	\hat{\mu}_{AIPWT}=\frac{1}{n}\sum_{k=1}^n\{\frac{R_kY_k}{Z_k}-\frac{R_k-Z_k}{Z_k}m(\mathbf{X}_k,\hat{\beta})\}
\end{equation}
where $m(\mathbf{X}_k,\hat{\beta})$ is the conditional mean of $Y_k$ and $Z_k$ is the conditional propensity of response. Typically the conditional mean of $Y_k$ is estimated by multiple linear regression (MLR)
\begin{equation}
	\label{rrz_mean}
	m(\mathbf{X}_k,\hat{\beta})=E[Y_k|\mathbf{X}_k;\hat{\beta}]=\hat{\beta}_0+\hat{\beta}_1X_{k1}+\ldots+\hat{\beta}_pX_{kp}.
\end{equation}
For $Z_k$, logistic regression is typically used,
\begin{equation}
	\label{prop_freq}
	Z_k=P(R_k=1|\mathbf{X}_k)=\frac{\exp(\mathbf{X}_k\hat{\mathbf{\theta})}}{1+\exp(\mathbf{X}_k\hat{\mathbf{\theta}})}.
\end{equation}
For implementation ease, we may re-write equation (\ref{rrz}) as
\begin{equation*}
	\hat{\mu}_{AIPWT}=\frac{1}{n}\sum_{k=1}^n\{\frac{R_k}{Z_k}[Y_k-m(\mathbf{X}_k,\hat{\mathbf{\beta}})]+m(\mathbf{X}_k,\hat{\mathbf{\beta}})\},
\end{equation*}
the sum of weighted mean of the MLR residuals for observed $Y_k$ and the mean of the predicted $Y_k$s under MLR for the missing $Y_k$s. A recap of the consistency of the AIPWT estimator is given in Web Appendix A. Briefly, the AIPWT estimator is doubly robust because we may write the expectation of $\hat{\mu}_{AIPWT}$ as $\mu$ plus the expectation of the propensity model multiplied by the mean model. Under MAR assumption, the multiplication of the propensity and mean model may be separated and hence, either the correct specification of the mean or propensity model would result in the expectation of the propensity model multiplied by the mean model to be 0.

\noindent
{\bf{2.3 Penalized splines of propensity prediction (PSPP)}}.

Another commonly used double robust estimator for the population mean $\mu$ is the PSPP model \citep{zhang_little}. This model proceeds by computing equation (\ref{prop_freq}) followed by imputing $Y_k$ using
\begin{equation}
	\label{pspp}
	Y_k=s[Z_k|\phi]+f(X_{k1},\ldots,X_{kp}),\eta)+\epsilon_k
\end{equation}
where $\epsilon_k\sim N(0,\sigma^2)$ and $s[Z_k|\phi]$ is the penalized spline formulation with $H$ fixed knots for $Z_k$ \citep{ruppert} and $f(X_{k1},\ldots,X_{kp}),\eta)$ is any arbitrary function with $\eta$ representing the parameters in the function $f(.)$. The common recommendation for the function $f(.)$ is a linear regression model. For $s[P(Z_k)|\phi]$, we consider a penalized linear mixed effect model using cubic splines.
%a mixed effects model of the form $\phi_0+\sum_{l=1}^L\phi_lZ_k^l+\sum_{h=1}^H\phi_{L+h}(Z_k-\tau_h)_+^L$ where $\phi_0,..,\phi_L$ are fixed parameters and $\phi_{L+1},..,\phi_{L+H}$ are random effects distributed iid $N(0,\sigma^2_{\phi})$. Assuming a linear form for $g(.)$, the predicted values of $y$ are given by $\hat{y}=C(C^TC+\hat{\lambda}D)^{-1}C^Ty$, where $C=[C_1 \: C_2]$ for the fixed effects matrix $C_1=[1,Z,..,Z^L,x_1,..,x_p]$ and the random effect matrix $C_2=[(Z-\tau_1)^L,...,(Z-\tau_H)^L]$, $D=diag(0_{L+p+1},I_H)$, and $\hat{\lambda}=\hat{\sigma}^2/\hat{\sigma}^2_{\phi}$ for the REML estimates of $\sigma^2$ and $\sigma^2_{\phi}$. Here we consider the penalized cubic spline models ($L=3$).
Alternative spline basis matricies are possible, as are alternative fitting methods, e.g., via direct minimization of the penalty function via cross-validation.

A review of the double robustness properties of the PSPP estimator is given in Web Appendix B. Briefly, when the mean model is specified correctly, the propensity model may be treated as random noise and hence the PSPP estimate is consistent for $\mu$. Now suppose that the propensity model is specified correctly, and we omit the mean model. By the balancing property of the propensity score, $E[Y_k|Z_k]=g(Z_k)$ for an unknown function $g(.)$. Using a cubic spline model for $g(.)$ allows the robust estimation of $g(Z_k)$ i.e., $E[Y_k|Z_k]=g(Z_k)\overset{p}{\rightarrow}\mu$. \cite{zhang_little} showed that this property can be extended to any misspecified form of $f(X_{k1},\ldots,X_{kp},\eta)$ so that $E[Y_k|Z_k,\mathbf{X}_k]=g(Z_k,\mathbf{X}_k)\overset{p}{\rightarrow}\mu$ if the propensity model is correctly specified.

\noindent
{\bf{3. Proposed methods}}.

We begin with a brief description of Bayesian additive regression trees (BART) before describing the use of BART in the various missing data approaches.

\noindent
{\bf{3.1 Bayesian additive regression trees}}

\noindent
{\emph{3.1.1 Continuous outcomes}}.
Suppose a continuous outcome $Y_k$ with associated $p$ covariates $\mathbf{X}_k=(X_{k1},\ldots,X_{kp})^T$ for $k=1,\ldots,n$ subjects. BART models the outcome as
\begin{equation}
	\label{bart_main_con_c2}
	Y_k=\sum_{j=1}^m g(\mathbf{X}_k,T_j,\mathbf{M}_j)+\epsilon_k \quad\epsilon_k\overset{i.i.d.}{\sim} N(0,\sigma^2)
\end{equation}
where $T_j$ is the $j^{\text{th}}$ binary tree structure and $\mathbf{M}_j=(\mu_{1j},\ldots,\mu_{b_jj})^T$ is the set of $b_j$ terminal node parameters associated with tree structure $T_j$ \citep{chipman_bart}. The function $g(\mathbf{X}_k,T_j,\mathbf{M}_j)$ can be viewed as the $j^{\text{th}}$ function that assigns the mean $\mu_{ij}$ to the $k^{\text{th}}$ outcome, $Y_k$. Typically, the number of trees $m$ is fixed and no prior distribution is placed on $m$. Chipman et al. suggested setting $m=200$ as this performs well in many situations. Alternatively, cross-validation could be used to determine $m$ \citep{chipman_bart}.

The joint prior distribution for (\ref{bart_main_con_c2}) is
\begin{equation}
	\label{bart_prior}
	P[(T_1,\mathbf{M}_1),\ldots,(T_m,\mathbf{M}_m),\sigma].
\end{equation}
Assuming $\epsilon_k$ and $(T_j,\mathbf{M}_j)$ are independent and all $m$ tree structures and terminal node parameters are independent between each other, we decompose equation (\ref{bart_prior}) to become
\begin{equation}
	\label{decom_prior}
	[\prod_{j=1}^m\{\prod_{i=1}^{b_j}P(\mu_{ij}|T_j)\}P(T_j)]P(\sigma)
\end{equation}
where $i=1,\ldots,b_j$ indexes the terminal node parameters in tree $j$. Assigning priors to $T_j$, $\mu_{ij}|T_j$, and $\sigma$ completes the BART model. The posterior draw of
$P[(T_1,\mathbf{M}_1),\ldots,(T_m,\mathbf{M}_m),\sigma|Y_k]$ is achieved using a combination of Bayesian backfitting \citep{hastie} and Metropolis within Gibbs algorithm. Details of the suggested priors and hyperparameters for $T_j$, $\mu_{ij}|T_j$, and $\sigma$ as well as the Bayesian backfitting and Metropolis within Gibbs algorithm can be found in \cite{chipman_bart}.

\noindent
{\emph{3.1.2 Binary outcomes}}.
Extending BART to binary outcomes involve a modification of (\ref{bart_main_con_c2}). First, let
\begin{equation}
	\label{bart_main_bin_c2}
	G(\mathbf{X}_k)=\sum_{j=1}^m g(\mathbf{X}_k,T_j,\mathbf{M}_j).
\end{equation}
Using the probit formulation, the binary outcomes $Y_k$ can be linked to (\ref{bart_main_bin_c2}) using $P(Y_k=1|\mathbf{X}_k)=\Phi[G(\mathbf{X}_k)]$ where $\Phi[.]$ is the cumulative density function of a standard normal distribution. This formulation implicitly assumes that $\sigma\equiv 1$. Assuming once again that all $m$ tree structures and terminal node parameters are independent, this implies that we only need priors for $T_j$ and $\mu_{ij}|T_j$. Further details regarding the prior distribution of binary outcomes BART can be found in \cite{chipman_bart}. To draw from the posterior distribution, Chipman et al. proposed the use of data augmentation \citep{albert,tanner}. This method proceeds by first generating a latent variable $Z_k$ according to
\begin{align*}
	(Z_k|Y_k=1,\mathbf{X}_k)&\sim N_{(0,\infty)}(G(\mathbf{X}_k),1)\\
	(Z_k|Y_k=0,\mathbf{X}_k)&\sim N_{(-\infty,0)}(G(\mathbf{X}_k),1),
\end{align*}
where $N_{(a,b)}(\mu,\sigma^2)$ is the truncated normal distribution with mean $\mu$ and variance $\sigma^2$ truncated to the range $(a,b)$. Once $Z_k$ is drawn, it is used to replace $Y_k$ in the algorithm to calculate the posterior distribution of continuous outcomes BART with $\sigma$ fixed at 1. Note that at each iteration, $G(\mathbf{X}_k)$ will be updated with the new $(T_1,\mathbf{M}_1),\ldots,(T_m,\mathbf{M}_m)$ draws from $P[(T_1,\mathbf{M}_1),\ldots,(T_m,\mathbf{M}_m)|Z_k]$ so that an updated draw of the latent variable $Z_k$ can be obtained.

The main advantage of BART lies in its ability to incorporate non-linear main effects and non-linear multiple-way interactions easily without any specification from the researcher. The main effects and multiple-way interactions are taken care of by BART because the Metropolis within Gibbs algorithm allows each regression tree to grow and change in order to explore a small portion of the outcome space. As each regression tree grows and changes, internal nodes where the parent and children nodes involve splitting on only one variable will constitute a main effect, while internal nodes where the parent and children nodes involve splitting on different variables would imply an interaction between these variables. When the regression trees are summed together, the different regression tree structures produce a non-linear estimation of both the main and multiple-way interaction effects.

\noindent
{\bf{3.2 Modifying the augmented inverse probability estimator with BART}}.

To modify the AIPWT, we replace $Z_k$ in equation (\ref{rrz}) with $Z_k^*$ where
\begin{equation}
	\label{bart_freq}
	Z_k^*=P(R_k=1|\mathbf{X}_k)=\Phi[G(\mathbf{X}_k)],
\end{equation}
with $G(\mathbf{X}_k)$ estimated using equation (\ref{bart_main_bin_c2}). Next, we model $m(\mathbf{X}_k,\hat{\beta})$ as a sum of regression trees i.e replace equation (\ref{rrz_mean}) with
\begin{equation}
	\label{rrz_mean_bart}
	Y_k=\sum_{j=1}^m g(\mathbf{X}_k,T_j,\mathbf{M}_j)+\epsilon_k,
\end{equation}
where $\epsilon_k\overset{\text{i.i.d.}}{\sim}N(0,\sigma^2)$. This would allow the propensity model and mean model to be approximately close to the true generating model if the model contains non-linear main and/or multiple-way interaction effects. BART will converge to the underlying true function of the propensity and mean models as long as the missingness mechanism is MAR and all of the variables that predict the propensity model or mean model are included in BART \citep{rockova}.

\noindent
{\bf{3.3 Modifying PSPP using BART: Penalized splines of BART propensity prediction (PSBPP)}}.

For PSPP, we modify it by replacing the estimation of $Z_k$ in equation (\ref{pspp}) with BART i.e. replace $Z_k$ with $Z_k^*$. This gives
\begin{equation}
	\label{psbpp}
	Y_k=\phi_0+\sum_{l=1}^L\phi_lZ_k^{*l}+\sum_{h=1}^H\phi_{L+h}(Z_k^*-\tau_h)_+^L+f(X_{k1},\ldots,X_{kp}),\eta)+\epsilon_k.
\end{equation}
Since BART was used to estimate the propensity score, we call this the penalized splines of BART propensity prediction (PSBPP).

\noindent
{\bf{3.4 Imputing directly using BART}}.

In \cite{ks}, they argued that using the mean model is more appropriate in situations where misspecifying both the propensity and mean model is high. Since BART has the potential to approximate models with non-linear main and multiple-way interaction effects closely, it may be more straight forward to impute $Y_k$ directly using BART. This idea is not new and has been used previously by \cite{xu} in a sequential fashion as we have pointed out in our introduction. In essence, the missing $Y_k$ outcomes are imputed using equation (\ref{rrz_mean_bart}).

\noindent
{\bf{3.5 Adding the BART propensity score to BART (BARTps)}}.

Although PSPP uses a spline to reduce model misspecification for the prediction of $Y_k$ given $Z_k$, possible interaction with $\mathbf{X}_k$ might still be present. Hence, using BART at both stages of modeling may be worth considering where
\begin{equation}
	\label{bartps}
	Y_k=\sum_{j=1}^m g(Z_k^*,\mathbf{X}_k,T_j,\mathbf{M}_j)+\epsilon_k,
\end{equation}
with $\epsilon_k\overset{i.i.d.}{\sim}N(0,\sigma^2)$, i.e. impute the missing $Y_k$ outcomes using equation (\ref{rrz_mean_bart}) with the addition of the BART estimated propensity score $Z_k^*$ as a predictor.

\noindent
{\bf{4. Simulations}}.
\label{sim}

Below we describe the three different simulation scenarios we used to investigate how misspecification due to incorrect model would affect the performance of PSPP, AIPWT, PSBPP, AIPWT with BART, BART, and BARTps. For reference, we included the usual sample mean estimator before partial removal of outcomes, labeled as BD, the complete case estimation of the sample mean, labeled as CC, as well as imputation using only the mean model, labeled as MLR.

For each of the simulation scenarios, we further split them into four situations: 1. both the propensity and mean models are correctly specified; 2. the mean model is misspecified but the propensity model is correctly specified; 3. the propensity model is misspecified but the mean model is correctly specified; and 4. both models are misspecified. For every method under the four situations and three scenarios, 500 simulations were used to estimate the empirical bias, root mean squared error (RMSE), 95\% coverage, and average length of the 95\% confidence interval (AIL). For PSPP and PSBPP, we used the linear truncated basis with 20 equally spaced knots on the propensity score, $Z_k$ or $Z_k^*$, to estimate the penalized splines. We estimated the penalized splines following the method described in Chapter 9 of \cite{ruppert}. The 95\% confidence interval (CI) and the length of this interval were estimated using a modified bootstrap approach with 200 resamples \citep{heitjan}. In brief, the modified bootstrap approach proposed by \cite{heitjan} resamples the data $D$ times calculating the desired estimate $\hat{y}_d$ for each resampled dataset. Then, Rubin's combining rule is applied to $\hat{y}_d$ to obtain the estimate and uncertainty. This modified bootstrap approach accounts for the uncertainty of the parameter estimates in the imputation model. In addition to bootstrap, we also performed MI using the posterior mean of the propensity score in equations (\ref{pspp}), (\ref{psbpp}), and (\ref{bartps}) as well as MI using a posterior draw of the propensity score in equations (\ref{pspp}), (\ref{psbpp}), and (\ref{bartps}). Finally, we considered sample sizes of 500, 1,000, and 5,000 to investigate how changes in sample size would affect the performance of each estimator.

\label{linear_mean_model}
\noindent
{\bf{4.1 Linear interaction in mean model}}

For this scenario, we included a linear two-way interaction term in both the propensity and mean model. We first generated 2 predictors as $X_{k1}\sim N(0,0.5)$ and $X_{k2}=X_{k1}+W_k$ where $W_k\sim N(0.25,0.5)$. We then specified the true propensity model as
\begin{equation}
		\label{prop_model}
		\text{logit}[P(M_k=1|X_{k1},X_{k2})]=\frac{1}{3}\{0.15+0.75(X_{k1}+X_{k2})-2X_{k1}X_{k2}\}
\end{equation}
and the mean model as
\begin{equation}
		\label{mean_model_linear}
		Y_k=10.8125+0.75(X_{k1}+X_{k2})-2X_{k1}X_{k2}+\epsilon_k
\end{equation}
where $\epsilon_k\overset{\text{iid}}{\sim}N(0,2^2)$. The resulting population mean for this model is 10.

We consider four types of model misspecification:
\begin{enumerate}
	\item[(i)] Propensity model and mean model are specified correctly as equations (\ref{prop_model}) and (\ref{mean_model_linear}),
	\item[(ii)] Mean model is misspecified by dropping the interaction term in equation (\ref{mean_model_linear}),
	\item[(iii)] Propensity model is misspecified by dropping the interaction term in equation (\ref{prop_model}), and
	\item[(iv)] Both propensity and mean models are misspecified by dropping the interaction terms in equations (\ref{prop_model}) and (\ref{mean_model_linear}).
\end{enumerate}

For BD and CC, note that because these estimators do not involve the specification of a propensity or mean model when estimating the population parameter $\mu$, the estimators will be the same under all situations. For MLR, since it does not involve the specification of a propensity model, the MLR estimate under situations (i) and (iii), and (ii) and (iv) will be the same. Because BART automatically takes care of non-linear main effects and non-linear multiple-way interaction effects, the PSBPP estimator under situations (i) and (iii), and (ii) and (iv) will be the same.  For the AIPWT with BART, BART, and BARTps methods, because each of them rely on BART to estimate their propensity and mean model, the estimators for all four situations will be the same.

\noindent
{\bf{4.2 Quadratic interaction in mean model}}

We further explore a slightly more complicated scenario with the propensity model still being the same i.e. equation (\ref{prop_model}) for the propensity model, but the mean model is now
\begin{equation}
		\label{mean_model_quad}
		Y_k=11.875+0.75(X_{k1}+X_{k2})-2(X_{k1}X_{k2})^2+\epsilon_k
\end{equation}
where $\epsilon_k\overset{\text{iid}}{\sim}N(0,2^2)$. $X_{k1}$ and $X_{k2}$ are generated as in subsection 4.1.1 and the population mean for this model is still 10. This scenario allows us to see how a slight non-linear effect in the simple two-way interaction of the mean model would affect the results of the eight mean estimation methods. The misspecification of the four situations is similar to the previous section in that the misspecification will remove the two-way interaction term.

\label{kseg}
\noindent
{\bf{4.3 Kang and Schafer (2007) example}}

In our third scenario, we explored how our proposed methods would perform under the scenario proposed by \cite{ks}, which we describe here. The propensity model is given by \begin{equation}
		\label{prop_model_ks}
		\text{logit}[P(R_k=1|U_{k1},U_{k2},U_{k3},U_{k4})]=-U_{k1}+0.5U_{k2}-0.25U_{k3}-0.1U_{k4},
\end{equation}
where $U_{kj}\overset{\text{iid}}{\sim}N(0,1)$, $j=1,\ldots,4$. The mean model is given by
\begin{equation}
		\label{mean_model_ks}
		Y_k=210+27.4U_{k1}+13.7(U_{k2}+U_{k3}+U_{k4})+\epsilon_k
\end{equation}
where $\epsilon_k\overset{\text{iid}}{\sim}N(0,1)$. In the misspecification situations, we assume that the $U_{kj}$s are latent and we only observe $X_{kj}$s which are given by
\begin{align*}
	X_{k1}&=\frac{\exp[U_{k1}]}{2},\\
	X_{k2}&=\frac{U_{k2}}{1+\exp[U_{k1}]},\\
	X_{k3}&=[\frac{U_{k1}U_{k3}}{25}+0.6]^3,\text{ and}\\
	X_{k4}&=[U_{k2}+U_{k4}+20]^2.
\end{align*}

For the four situations, we use $U_{kj}$s to estimate the propensity and mean model when both models are specified correctly. When the propensity model is specified correctly but the mean model is misspecified, we use $U_{kj}$ to estimate the propensity model but replace the $U_{kj}$ with $X_{kj}$ when estimating the mean model. When the mean model is specified correctly but the propensity model is misspecified, we replace $U_{kj}$ with $X_{kj}$ to estimate the propensity model but use $U_{kj}$ to estimate the mean model. When both propensity and mean model are misspecified, we replace $U_{kj}$ with $X_{kj}$s to estimate both the propensity and mean model.

\noindent
{\bf{4.4 Results}}

\begin{table}
\caption{Bias, RMSE, 95\% coverage, and average 95\% confidence interval length (AIL) of the eight estimators under the linear interaction in mean model scenario with sample size 1,000.}
\label{lin_res_1000}
\begin{center}
	\hspace*{-1cm}
	\begin{tabular}{|lcccc|cccc|}
		\hline
		$n=1,000$ & \multicolumn{4}{c|}{Both correct} & \multicolumn{4}{c|}{Propensity correct} \\
		Method & Bias & RMSE & Coverage & AIL & Bias & RMSE & Coverage & AIL \\ \hline
		BD	&	0	&	0.09	&	94.2	&	0.34	&	0	&	0.09	&	94.2	&	0.34	\\
CC	&	0.51	&	0.53	&	0.6	&	0.42	&	0.51	&	0.53	&	0.6	&	0.42	\\
MLR	&	0	&	0.12	&	99	&	0.62	&	0.45	&	0.46	&	10	&	0.57	\\
PSPP	&	0.01	&	0.14	&	99.8	&	0.78	&	0.05	&	0.13	&	97.4	&	0.61	\\
AIPWT	&	0	&	0.12	&	94.4	&	0.47	&	0.04	&	0.18	&	87.2	&	0.6	\\
PSBPP	&	0	&	0.13	&	99.2	&	0.64	&	-0.06	&	0.15	&	98.4	&	0.71	\\
AIPWT with BART	&	0.11	&	0.17	&	78.2	&	0.44	&	0.11	&	0.17	&	78.2	&	0.44	\\
BART	&	0.14	&	0.19	&	87.4	&	0.57	&	0.14	&	0.19	&	87.4	&	0.57	\\
BARTps	&	0.07	&	0.14	&	95.8	&	0.6	&	0.07	&	0.14	&	95.8	&	0.6	\\
		\hline
		& \multicolumn{4}{c|}{Mean correct} & \multicolumn{4}{c|}{Both wrong}\\
		Method & Bias & RMSE & Coverage & AIL & Bias & RMSE & Coverage & AIL \\
		\hline
		BD	&	0	&	0.09	&	94.2	&	0.34	&	0	&	0.09	&	94.2	&	0.34	\\
CC	&	0.51	&	0.53	&	0.6	&	0.42	&	0.51	&	0.53	&	0.6	&	0.42	\\
MLR	&	0	&	0.12	&	99	&	0.62	&	0.45	&	0.46	&	10	&	0.57	\\
PSPP	&	0	&	0.12	&	99	&	0.63	&	0.22	&	0.26	&	85	&	0.78	\\
AIPWT	&	0	&	0.12	&	92.8	&	0.46	&	0.43	&	0.45	&	5	&	0.43	\\
PSBPP	&	0	&	0.13	&	99.2	&	0.64	&	-0.06	&	0.15	&	98.4	&	0.71	\\
AIPWT with BART	&	0.11	&	0.17	&	78.2	&	0.44	&	0.11	&	0.17	&	78.2	&	0.44	\\
BART	&	0.14	&	0.19	&	87.4	&	0.57	&	0.14	&	0.19	&	87.4	&	0.57	\\
BARTps	&	0.07	&	0.14	&	95.8	&	0.6	&	0.07	&	0.14	&	95.8	&	0.6	\\
		\hline
	\end{tabular}
\end{center}
\end{table}

Table \ref{lin_res_1000} shows the result under the scenario of a simple linear interaction term in the mean model for a sample size of 1,000. The CC estimators were substantially biased under all four types of misspecification. When the propensity model was correctly specified, both PSPP and AIPWT were approximately unbiased, although PSPP had much smaller RMSE and better coverage. The MLR, PSPP, and AIPWT estimators performed very well in terms of bias and RMSE when the mean model was correctly specified. When both models were misspecified, MLR, PSPP, and AIPWT were biased with coverage of both models decreasing dramatically. For PSBPP and AIPWT with BART, we observed that specifying the propensity model of PSPP using BART had little effect on the bias, RMSE, 95\% coverage, and AIL when either one or both the propensity and mean model were correctly specified. When both models were misspecified, PSBPP was still able to produce nearly unbiased estimation of the population mean and relatively similar AIL as the propensity model used can be considered correct. In contrast, AIPWT with BART had bias and relatively poor coverage compared to AIPWT when at least one of the models in AIPWT was specified correctly. AIPWT with BART only performed better than AIPWT when both models were misspecified. Still, some bias and below nominal coverage remained. AIPWT with BART was more biased with larger RMSE and poorer coverage compared to PSBPP under all scenarios. The BART estimators alone generally had performance similar to AIPWT with BART, if slightly poorer in terms of bias and RMSE. For BARTps, the bias was reduced compared to a BART model with only $X_k$s as the predictors. Addition of BART propensity scores $Z_k^*$ also improves the 95\% coverage compared to BART; nominal coverage was achieved for BARTps.

As the sample size increases, the bias, RMSE, and AIL of all methods reduce, and nominal 95\% coverage increases (See Tables \ref{app_lin_500} to \ref{app_lin_5000} in Web Appendix C). MI results were similar to bootstrap results (See Tables \ref{app_lin_500_mi} to \ref{app_lin_5000_mi} in Web Appendix C). Using a posterior draw of the propensity scores instead of posterior mean increased bias slightly for PSBPP and BARTps (See Tables \ref{app_lin_500_mi_new} to \ref{app_lin_5000_mi_new} in Web Appendix C).

\begin{table}
\caption{Bias, RMSE, 95\% coverage, and average 95\% confidence interval length (AIL) of the eight estimators under the quadratic interaction in mean model scenario with sample size 1,000.}
\label{quad_res_1000}
\begin{center}
	\hspace*{-1cm}
	\begin{tabular}{|lcccc|cccc|}
		\hline
		$n=1,000$ & \multicolumn{4}{c|}{Both correct} & \multicolumn{4}{c|}{Propensity correct} \\
		Method & Bias & RMSE & Coverage & AIL & Bias & RMSE & Coverage & AIL \\ \hline
		BD	&	0	&	0.24	&	91.8	&	0.86	&	0	&	0.24	&	91.8	&	0.86	\\
CC	&	1.21	&	1.23	&	0.2	&	0.63	&	1.21	&	1.23	&	0.2	&	0.63	\\
MLR	&	0	&	0.26	&	99	&	1.32	&	1.24	&	1.25	&	0	&	0.8	\\
PSPP	&	0	&	0.26	&	98.8	&	1.33	&	0.21	&	0.44	&	81.2	&	2	\\
AIPWT	&	0	&	0.26	&	91.2	&	0.93	&	0.22	&	0.72	&	67	&	1.68	\\
PSBPP	&	0	&	0.26	&	98.6	&	1.33	&	0.13	&	0.35	&	94	&	2.16	\\
AIPWT with BART	&	0.45	&	0.51	&	29.8	&	0.77	&	0.45	&	0.51	&	29.8	&	0.77	\\
BART	&	0.52	&	0.57	&	42	&	0.97	&	0.52	&	0.57	&	42	&	0.97	\\
BARTps	&	0.41	&	0.47	&	63.4	&	1.07	&	0.41	&	0.47	&	63.4	&	1.07	\\
		\hline
		& \multicolumn{4}{c|}{Mean correct} & \multicolumn{4}{c|}{Both wrong}\\
		Method & Bias & RMSE & Coverage & AIL & Bias & RMSE & Coverage & AIL \\
		\hline
		BD	&	0	&	0.24	&	91.8	&	0.86	&	0	&	0.24	&	91.8	&	0.86	\\
CC	&	1.21	&	1.23	&	0.2	&	0.63	&	1.21	&	1.23	&	0.2	&	0.63	\\
MLR	&	0	&	0.26	&	99	&	1.32	&	1.24	&	1.25	&	0	&	0.8	\\
PSPP	&	0	&	0.26	&	98.6	&	1.33	&	0.72	&	0.77	&	61.8	&	1.69	\\
AIPWT	&	0	&	0.25	&	91	&	0.92	&	1.21	&	1.22	&	0	&	0.59	\\
PSBPP	&	0	&	0.26	&	98.6	&	1.33	&	0.13	&	0.35	&	94	&	2.16	\\
AIPWT with BART	&	0.45	&	0.51	&	29.8	&	0.77	&	0.45	&	0.51	&	29.8	&	0.77	\\
BART	&	0.52	&	0.57	&	42	&	0.97	&	0.52	&	0.57	&	42	&	0.97	\\
BARTps	&	0.41	&	0.47	&	63.4	&	1.07	&	0.41	&	0.47	&	63.4	&	1.07	\\
		\hline
	\end{tabular}
\end{center}
\end{table}

Table \ref{quad_res_1000} shows the result under the scenario of a quadratic interaction term in the mean model for a sample size of 1,000. This scenario was more challenging compared to the linear interaction term scenario, with larger bias, RMSE, and AIL with smaller 95\% coverage for all methods. For the PSPP and AIPWT method, when the propensity model was correctly specified or when the mean model was correctly specified, we started to see substantial increases in the bias, RMSE, and AIL with a substantial reduction in the 95\% coverage. When both models were misspecified, we started to see very poor performance: bias, RMSE, and AIL further increased with further reduction in the 95\% coverage. For the PSBPP, the bias, RMSE, and AIL were similar to PSPP when either both models were correctly specified or only one model was correctly specified, although when the mean model was misspecified, PSBPP produced a better nominal 95\% coverage. When both models were misspecified, PSBPP performed the best compared to all the other six estimators with modest bias and approximately correct nominal coverage. For AIPWT with BART, once again BART was able to help the AIPWT estimator when both propensity and mean models were misspecified but when either one or both models were correctly specified, AIPWT with BART performed worse compared to AIPWT. In addition, the performance of AIPWT with BART when both propensity and mean models were misspecified was not as good compared to PSBPP. For BART, we noted similar conclusions as the linear interaction in mean model scenario with similar performance to AIPWT with BART. Adding BART propensity scores $Z_k^*$ as a predictor reduces the bias and RMSE and improves the 95\% coverage slightly compared to BART, but bias remains and nominal coverage is still somewhat poor.

Similar to the linear interaction in mean model scenario, we found that as sample size increases, the bias, RMSE, and AIL of all methods reduce while 95\% coverage increases (See Tables \ref{app_quad_500} to \ref{app_quad_5000} in Web Appendix C). MI results echo those observed using bootstrap (See Tables \ref{app_quad_500_mi} to \ref{app_quad_5000_mi} in Web Appendix C) while MI results using a posterior draw of the propensity score once again produced an increase in bias for PSBPP and BARTps methods (See Tables \ref{app_quad_500_mi_new} to \ref{app_quad_5000_mi_new} in Web Appendix C).

\begin{table}
\caption{Bias, RMSE, 95\% coverage, and average 95\% confidence interval length (AIL) under the Kang and Schafer (2007) example with sample size 1,000.}
\label{ks_res_1000}
\begin{center}
	%\tiny
	\hspace*{-1cm}
	\begin{tabular}{|lcccc|cccc|}
		\hline
		$n=1,000$ & \multicolumn{4}{c|}{Both correct} & \multicolumn{4}{c|}{Propensity correct} \\
		Method & Bias & RMSE & Coverage & AIL & Bias & RMSE & Coverage & AIL \\ \hline
		BD	&	0.07	&	1.11	&	95.2	&	4.38	&	0.07	&	1.11	&	95.2	&	4.38	\\
CC	&	-9.96	&	10.09	&	0	&	5.97	&	-9.96	&	10.09	&	0	&	5.97	\\
MLR	&	0.07	&	1.11	&	99.4	&	6.38	&	-0.74	&	1.63	&	98	&	7.78	\\
PSPP	&	0.06	&	1.11	&	99.4	&	6.38	&	-0.07	&	1.21	&	99.2	&	6.66	\\
AIPWT	&	0.06	&	1.11	&	95.6	&	4.38	&	0.07	&	1.66	&	94.2	&	6.01	\\
PSBPP	&	0.07	&	1.11	&	99.4	&	6.38	&	1.46	&	1.95	&	96.8	&	7.4	\\
AIPWT with BART	&	-0.05	&	1.12	&	95.2	&	4.42	&	-0.31	&	1.19	&	93.8	&	4.61	\\
BART	&	-0.13	&	1.12	&	99.6	&	6.38	&	-0.59	&	1.29	&	99.2	&	6.5	\\
BARTps	&	0	&	1.11	&	99.4	&	6.46	&	0.39	&	1.23	&	99.2	&	6.8	\\
		\hline
		& \multicolumn{4}{c|}{Mean correct} & \multicolumn{4}{c|}{Both wrong}\\
		Method & Bias & RMSE & Coverage & AIL & Bias & RMSE & Coverage & AIL \\	\hline
		BD	&	0.07	&	1.11	&	95.2	&	4.38	&	0.07	&	1.11	&	95.2	&	4.38	\\
CC	&	-9.96	&	10.09	&	0	&	5.97	&	-9.96	&	10.09	&	0	&	5.97	\\
MLR	&	0.07	&	1.11	&	99.4	&	6.38	&	-0.74	&	1.63	&	98	&	7.78	\\
PSPP	&	0.07	&	1.11	&	99.4	&	6.38	&	-2.12	&	2.52	&	77.2	&	6.29	\\
AIPWT	&	-0.08	&	2.28	&	95.6	&	5.1	&	-35.69	&	477.13	&	41.2	&	196.51	\\
PSBPP	&	0.07	&	1.11	&	99.4	&	6.38	&	-1.13	&	1.73	&	99	&	7.84	\\
AIPWT with BART	&	-0.06	&	1.12	&	95.2	&	4.42	&	-0.45	&	1.24	&	93.2	&	4.62	\\
BART	&	-0.13	&	1.12	&	99.6	&	6.38	&	-0.59	&	1.29	&	99.2	&	6.5	\\
BARTps	&	-0.05	&	1.12	&	99.6	&	6.46	&	-0.52	&	1.27	&	99.2	&	6.7	\\
		\hline
	\end{tabular}
\end{center}
\end{table}

Table \ref{ks_res_1000} shows the result under the \cite{ks} example for a sample size of 1,000. For the PSPP and AIPWT methods, we found once again that misspecification of the mean model increased the bias, RMSE, and AIL of these methods slightly more than misspecification of the propensity model does. When both models are misspecified, both models performed badly with the AIPWT estimator being highly unstable, producing a bias and RMSE more than the CC estimator. The standard MLR imputation performed fairly well even when the mean model was misspecified, as also reported by \cite{ks}. For the PSBPP and AIPWT with BART model, PSBPP performed better in terms of bias, RMSE, 95\% coverage, and AIL when both the propensity and mean models are correctly specified or when only the mean model is correctly specified. When only the propensity model is correctly specified or when both models are misspecified, PSPP and AIPWT with BART had similar (slightly below nominal) coverage; AIPWT with BART had reduced bias, RMSE, and smaller AIL. Compared to AIPWT and PSPP, AIPWT with BART and PSBPP respectively showed improvements in performance when both models are misspecified, with both having similar RMSE and slightly below nominal coverage. BART and BARTps generally performed well under all of the misspecification scenarios with BARTps having the better performance

Once again, we note that as sample size increases, the bias, RMSE, and AIL of all methods reduce (See Tables \ref{app_ks_500} to \ref{app_ks_5000} in Web Appendix C). The 95\% coverage of all methods remained relatively similar as the sample size increased except for PSPP and AIPWT where coverage decreased as sample size increased. MI results produced similar conclusions with bootstrap (See Tables \ref{app_ks_500_mi} to \ref{app_ks_5000_mi_new} in Web Appendix C).

%\noindent
%{\bf{4.3 Simulation summary}}.

%To summarize the results from all four simulation scenarios: PSBPP worked well when the non-linearity in the propensity and mean model involve only polynomial effects. However, when the main and interaction effects start to get complicated like the \cite{ks} example, BARTps worked better. With this conclusion in mind, we now investigate how PSBPP and BARTps perform when applied to the estimation of the population mean and various parameters from two real-life dataset.

\noindent
{\bf{5. Applications to Missing Data in Transportation Research}}.

\noindent
{\bf{5.1 Imputing Delta-v in 2014 National Automotive Sampling System Crashworthiness Data System dataset}}.

The NASS-CDS dataset is an annual three-stage representative probability sample of passenger vehicle crashes sponsored by the National Highway and Transportation Safety Authority (NHTSA). To be eligible for inclusion, a crash must: (1) be police reported, (2) involve a harmful event (property damage and/or personal injury) resulting from a crash, and (3) involve at least one towed passenger car or light truck or van in transport on a traffic way. When a crash is selected, NASS-CDS investigators obtain police reports and conduct interviews with the occupants to collect information such as driver’s age and sex, vehicle curb weight, type of vehicle, severity of injury measured using the KABCO scale \citep[K=fatal; A=incapacitating Injury; B=non-incapacitating injury; C=possible injury; O=no injury;][]{nhtsa}, and the principal direction of impact from the crash. Often, the variable that estimates instantaneous change in velocity (delta-v), is missing. This variable is important because many studies have shown that delta-v is a strong predictor for the severity of injuries in tow-away crashes.

The 2014 NASS-CDS dataset contains 3,660 non-rollover passenger vehicle crashes. We first converted all continuous variables to categorical. We then coded missingness in a variable as a level. We then removed variables that had more than 80\% missing, derived from other variables in the dataset, or 100\% missing for vehicles missing delta-v. Simple descriptive statistics of the variables in our dataset stratified by missingness in delta-v can be found in Tables \ref{des_cds1} to \ref{des_cds5} of Web Appendix D. Out of the 44 variables, only climate, body type of vehicle, whether the trajectory data was reconstructed, make of the vehicle, model year, number of occupants, pre-event movement, road alignment, road surface type, number of seriously injured occupants, and driver's age, height, and weight were \underline{not} statistically different between non-rollover passenger vehicles missing total delta-v and not missing delta-v.

We estimated the population mean of the 2014 total delta-v and the unadjusted odds ratio of the police reported injury severity (any injury or severe injury) as a function of delta-v (between 15kph and 35kph, and more than 35kph, versus less than 15kph).
We compared the estimate and 95\% confidence interval produced by CC, MLR, PSPP, AIPWT, PSBPP, and BARTps. To obtain the estimate and 95\% confidence interval for all six methods, we employed the finite Bayesian bootstrap method developed by \cite{zhou}. This procedure allows us to compute a valid estimate and 95\% confidence interval for our dataset while non-parametrically accouting for the sample design in the imputation.

\begin{table}
	\caption{Estimated population mean, and unadjusted odds ratios of injury severity where reference group is delta-v less than 15 kph ($X<15$). \label{anal_res_bbpp}}
	\scriptsize
	\hspace*{-2cm}
	\centering
	\begin{tabular}{|l|cr|cr|cr|cr|cr|}
		\hline		
			& \multicolumn{2}{c|}{\underline{$\bar{Y}_{\text{delta-v}}$}} & \multicolumn{2}{c|}{\underline{$OR_{\text{NULL}}$}} & \multicolumn{2}{c|}{\underline{$OR_{\text{NULL}}$}} & \multicolumn{2}{c|}{\underline{$OR_{\text{SEV}}$}} & \multicolumn{2}{c|}{\underline{$OR_{\text{SEV}}$}}\\	
		Method & Estimate & 95\% CI & $15\leq X\leq 35$ & 95\% CI & $X>35$ & 95\% CI & $15\leq X\leq 35$ & 95\% CI & $X>35$ & 95\% CI\\
		\hline
		CC	&	21.57	&	(	20.64	,	22.47	)	&	1.72	&	(	1.15	,	2.43	)	&	5.88	&	(	2.94	,	8.79	)	&	2.31	&	(	1.49	,	3.64	)	&	17.99	&	(	9.31	,	30.42	)	\\
MLR	&	21.57	&	(	20.64	,	22.47	)	&	1.37	&	(	1.08	,	1.76	)	&	2.78	&	(	1.92	,	3.52	)	&	1.43	&	(	1.19	,	1.69	)	&	6.08	&	(	4.25	,	8.17	)	\\
PSPP	&	21.55	&	(	20.06	,	22.99	)	&	1.86	&	(	1.23	,	2.84	)	&	7.93	&	(	3.66	,	13.33	)	&	3.19	&	(	1.82	,	5.21	)	&	33.73	&	(	16.16	,	60.17	)	\\
AIPWT	&	21.5	&	(	20.01	,	23.33	)	&	1.5	&	(	1	,	2.12	)	&	5.87	&	(	3.39	,	9.25	)	&	1.58	&	(	1	,	2.21	)	&	14.67	&	(	9.19	,	21.77	)	\\
PSBPP	&	21.75	&	(	18.29	,	25.61	)	&	1.86	&	(	1.11	,	3.03	)	&	7.67	&	(	2.39	,	13.74	)	&	3.3	&	(	1.51	,	6.8	)	&	33.23	&	(	9.63	,	71.67	)	\\
BARTps	&	21.9	&	(	18.51	,	24.79	)	&	1.62	&	(	1.08	,	2.36	)	&	2.95	&	(	1.54	,	5.18	)	&	1.77	&	(	1.18	,	2.64	)	&	7.48	&	(	4.09	,	12.27	)	\\
		\hline
	\end{tabular}
\end{table}

The result of our analysis is given in Table \ref{anal_res_bbpp}. The population mean of delta-v estimated by PSBPP and BARTps were similar, more than 21.6 kph while MLR, CC, PSPP, and AIPWT suggested that the population delta-v was about 21.5 kph. The 95\% confidence interval of PSBPP and BARTps were also slightly wider compared to MI, CC, PSPP, and AIPWT. For the odds ratios, PSPP and PSBPP tended to agree with each other under any injury, CC and AIPWT suggested somewhat similar results, while BARTps and MLR results were more similar. All methods suggested a significant association between delta-v and presence of injury with higher delta-v levels associated with a higher odds of experiencing injury in a non-rollover passenger vehicle crash. For severe versus non-severe injury, we observe similar results as injury versus no injury in that PSPP and PSBPP suggested similar results, CC and AIPWT suggested similar results, and BARTps and MLR suggested similar results. Again all methods suggested a significant association between delta-v and presence of injury with higher delta-v levels associated with a higher odds of experiencing injury in a non-rollover passenger vehicle crash. Given that CC results and AIPWT results were similar and BARTps and MLR results were similar, we suspect there to be non-linear main and interaction effects between delta-v and the NASS-CDS variables as well as non-linear main and interaction effects between the missingness of delta-v and the NASS-CDS variables.

\noindent
{\bf{5.2 Imputing Blood Alcohol Concentration levels in 2015 Fatality Analysis Reporting System dataset}}.

The FARS releases information annually from all fatal motor vehicle crashes that occur on US public roads. Information collected include person characteristics like age, gender, and BAC levels (g/100 ml); environmental conditions like weather, lighting, and surface conditions; vehicle characteristics like type of vehicle, model year of vehicle, and gross weight of vehicle; road characteristics like type of road, type of traffic control device, and alignment of road (straight, curved); and accident characteristics like accident type (rollover, collision), pre-accident maneuver, and time of accident. Of these information collected, BAC, which is used to identify alcohol involvement in fatal crashes, is often missing. The fact that alcohol involvement is more commonly reported in fatal crashes compared to personal injury and property-damage-only crashes makes this issue more concerning because high levels of missingness in BAC hinders the investigation of the trend and extent of alcohol involvement in fatal crashes, the successful identification of high-risk groups for counter measures, and evaluation of drunk-driving prevention programs.

Due to the importance of the BAC measure, NHTSA considered several approaches to remedy the missing data problem before deciding to use MI in 2002 \citep{subramaniam}. Although MI was a great improvement from previous imputation methods \citep{klein}, misspecification of the model in MI could lead to biased results. Replacing the imputation models with DR estimators like PSPP and AIPWT could further bias results if the propensity and mean model were not specified correctly. Hence, we applied our proposed methods to the 2015 FARS dataset to impute BAC levels and compared the imputation results with existing MI results provided by the FARS dataset.

Details of how the publicly available imputed BAC values for the 2015 dataset can be found in \cite{rubin} Section 3. We modified this imputation strategy slightly. First, we used the imputed 2015 BAC FARS dataset to determine all the 55,502 ``actively-involved'' subjects eligible for imputation (See Rubin et al. [1998], Section 2). We then restrict our attention to passenger vehicles as defined in Section 3 of Rubin et al. which gave us 19,425 subjects. We again recoded continuous variables as categorical variables and coded missing entries as a category in all variables. We then removed variables that had more than 80\% missing, derived from other variables in 2015 FARS, or 100\% missing for subjects missing BAC values. Simple descriptive statistics of the variables in our dataset stratified by missingness in BAC can be found in Tables \ref{des_fars1} to \ref{des_fars4} of Web Appendix D. All variables except whether crash occurred within the boundaries of a work zone were significantly different between subjects missing BAC and subjects not missing BAC.
We impute BAC values as follows:
\begin{enumerate}
	\item We first employed BART for binary outcomes to predict BAC=0 ($Y=0$) versus $BAC>0$ ($Y=1$) using all available predictors (See Tables \ref{des_fars1} to \ref{des_fars4} in Web Appendix D for all predictors employed).
	\item We set the predicted BAC=0 values as 0 and focus on the set of observed $BAC>0$ and predicted $BAC>0$. For the observed $BAC>0$, we employed a Box-Cox transformation \citep{boxcox} using all available predictors to obtain the Box-Cox transformation parameter $\hat{\lambda}$. We then used $\tilde{\lambda}=\hat{\lambda}+1$ as suggested by \cite{rubin}.
	\item We next imputed the Box-Cox transformed BAC value for the predicted $BAC>0$ using the following methods, PSPP, AIPWT, PSBPP, and BARTps. We only considered PSBPP and BARTps because both methods performed better in most of our simulation scenarios. For the transformed BAC values that were predicted to be negative, we set them as 0. For transformed BAC values that were predicted to be positive, an inverse transformation was applied to the predicted transformed BAC values to obtain the predicted BAC value in the original scale.
	\item We then drew 200 resampled datasets and repeated Steps 1-3 on each dataset to estimate the imputation uncertainty.
\end{enumerate}
For the estimate of interest, we examined the population mean of the BAC value, the proportion of BAC more than .010 g/100 ml, and the proportion of BAC more than .100 g/100 ml among passenger vehicles in 2015.

\begin{table}
	\caption{Estimated population mean of BAC, proportion of $BAC>.010$, and proportion of $BAC>.100$. All values in precentages.}
	\label{anal_res_bbpp2}
	\centering
	\begin{tabular}{|l|cr|cr|cr|}
		\hline
		& \multicolumn{2}{c|}{\underline{Mean}} & \multicolumn{2}{c|}{\underline{$BAC>1\%$}} & \multicolumn{2}{c|}{\underline{$BAC>10\%$}} \\
		Method & Estimate & 95\% CI & Estimate & 95\% CI & Estimate & 95\% CI\\
		\hline
		CC	&	5.72	&	(5.53, 5.91)	&	0.34	&	(0.33, 0.35)	&	0.26	&	(0.26, 0.27)	\\
MLR	&	3.97	&	(3.83, 4.11)	&	0.24	&	(0.24, 0.25)	&	0.18	&	(0.18, 0.19)	\\
PSPP	&	3.07	&	(2.89, 3.26)	&	0.18	&	(0.17, 0.19)	&	0.14	&	(0.13, 0.15)	\\
AIPWT	&	3.12	&	(2.10, 4.14)	&	0.16	&	(0.15, 0.16)	&	0.13	&	(0.13, 0.14)	\\
PSBPP	&	3.08	&	(2.88, 3.28)	&	0.18	&	(0.17, 0.19)	&	0.14	&	(0.13, 0.15)	\\
BARTps	&	3.13	&	(2.95, 3.27)	&	0.19	&	(0.18, 0.19)	&	0.15	&	(0.14, 0.15)	\\
		\hline
	\end{tabular}
\end{table}

Table \ref{anal_res_bbpp2} gives the result of our analysis. Results of the MLR method were calculated using the imputed BAC values provided in the 2015 FARS dataset. Comparing the results between CC and MLR, we can see that CC likely overestimates the population mean of BAC as well as the proportion of subjects with BAC more than .010 and .100 g/ 100 ml. MLR estimates that the population mean BAC value was 4\% with the proportion of subjects with BAC more than .010 estimated at 24\% and for the proportion of subjects with BAC more than .100 estimated at 18\%. The MLR results were significantly different from the imputed values estimated by PSPP and AIPWT. PSPP and AIPWT suggested that the population mean BAC value was about 3.1\% while the proportion of subjects with BAC more than .010 was estimated at about 18\% and 16\% respectively while the proportion of subjects with BAC more than .100 was estimated at about 14\%. PSBPP and BARTps results were similar compared to PSPP and AIPWT. The significant difference between MLR results versus the doubly robust and robust-squared methods suggest that there is likely some non-linear relation between BAC and the variables in the FARS dataset. The non-significant difference in the results produced by PSPP, PSBPP, and BARTps further suggests that the relationship between missingness in BAC and the rest of the FARS variables is linear without any interactions.

\noindent
{\bf{6. Discussion}}.

In many situations, researchers would not know the true propensity and mean model and thus both models have a high chance that they will be misspecified, limiting the value of the ``double-robustness'' property. Even if the misspecification was mild for example, removal of the two-way interaction terms when the true mean model included a linear two-way interaction term or quadratic two-way interaction term, the resulting bias may be almost as large as a complete case analysis. Hence we consider use of a highly flexible estimation model -- specifically Bayesian Additive Regression Trees or BART -- to reduce the risk of model misspecification. We consider the use of BART in propensity score estimation when using the penalized spline of propensity prediction (PSPPB) or when using the augmented inverse probability weighted estimator (AIPWT with BART). We also consider direct imputation using BART (BART), and a ``double flexible'' robust model that adds a BART-estimated propensity score to the BART imputation, so that both the mean and propensity are estimated in the PSPP model using BART (BARTps).

By using BART, we were able to demonstrate the reduction in bias and RMSE of the double robust estimators when both propensity and mean models were misspecified, with little loss in efficiency when either one or both of the mean and propensity models can be correctly specified by a standard linear or logistic model.
%We also show how BART could be used to improve the imputation results of the delta-v values in the 2015 NASS-CDS dataset as well as the BAC values in the 2015 FARS dataset. We recommend that when the true model is very complicated and misspecification of both the propensity and mean model is highly likely, using a very flexible model of BART either in PSPP, AIPWT, or directly greatly improves bias, RMSE, and coverage.
Our simulation study suggests that PSPP with BART performs considerably better than AIPWT with BART under settings with missing interaction terms. However, when both the propensity and mean model are complex, BARTps tends to perform better. Hence, we suggest PSBPP and BARTps as the preferred methods for imputing datasets under MAR, while acknowledging that these recommendations are empirically based on simulations that are somewhat limited in nature.

We also found in our simulation results that MI using a posterior draw of the propensity score in equations (\ref{psbpp}) and (\ref{bartps}) increased bias compared to using the posterior mean of the propensity score for linear and quadratic interaction scenarios. This is because the propensity model in both scenarios tended to create datasets where there is not much overlap in the predictors for response and non-response. Hence, the researcher might want to rely on bootstrap to obtain the uncertainty of PSBPP and BARTps during analysis.

Although we focused our attention on MAR for a continuous outcome, extension to a binary outcome is possible using generalized additive models or generalized linear mixed models for the PSPPB setting, or use of latent variables models (e.g, probit models) for PSPPB or the BARTps setting. The MAR assumption remains a restriction in these ``robust'' estimation methods; extensions to NMAR mechanisms remains a topic for further research.

\backmatter

%%%%%% include this section if you wish to acknowledge people,
%%%%%% grant support, etc.

%\noindent
%{\bf{Acknowledgments}}

%This work was supported jointly by Dr. Michael Elliott and Dr. Carol Flannagan.\vspace*{-8pt}

%%%%%% include this section only if your manuscript refers to supplementary
%%%%%% materials -- see Instructions for Authors at
%%%%%% http://www.tibs.org/biometrics

\noindent
{\bf{Supplementary Materials}}

Web Appendix A to D referenced throughout the paper are available
with this paper at the Biometrics website on Wiley Online Library.
\vspace*{-8pt}

\bibliographystyle{biom} \bibliography{references}

\appendix

\noindent
{\bf{Web Appendix A: Consistency of the AIPWT estimator}}

The AIPWT estimator is a consistent estimator for the population mean parameter $\mu$ when either the propensity model or mean model in equation (\ref{rrz}) is correctly specified. To see this, we first assume that $\hat{\mathbf{\beta}}\overset{p}{\rightarrow}\mathbf{\beta}^*$ and $\hat{\theta}\overset{p}{\rightarrow}\mathbf{\theta}^*$ i.e. the parameters in equations (\ref{rrz_mean}) and (\ref{prop_freq}) are consistent. This is valid since the models we used to estimate these parameters were multiple linear regression and multiple logistic regression which under the usual maximum likelihood assumptions, will converge asymptotically to their true values. From equation (\ref{rrz}), this implies that
\begin{align*}
	\hat{\mu}_{AIPWT}&=\frac{1}{n}\sum_{k=1}^n\{\frac{R_kY_k}{Z_k}-\frac{R_k-Z_k}{Z_k}m(\mathbf{X}_k,\hat{\beta})\}\\
	&\overset{p}{\rightarrow} E[\frac{R_kY_k}{Z_k}-\frac{R_k-Z_k}{Z_k}m(\mathbf{X}_k,\hat{\beta})]\\
	&=E[Y_k-Y_k+\frac{R_kY_k}{Z_k}-\frac{R_k-Z_k}{Z_k}m(\mathbf{X}_k,\hat{\beta})]\\
	&=\mu+E[\frac{R_kY_k}{Z_k}-\frac{R_k-Z_k}{Z_k}m(\mathbf{X}_k,\hat{\beta})-Y_k]\\
	&=\mu+E[\frac{R_k}{Z_k}Y_k-\frac{R_k}{Z_k}m(\mathbf{X}_k,\hat{\beta})-\{Y_k-m(\mathbf{X}_k,\hat{\beta})\}]\\
	&=\mu+E[\{\frac{R_k}{Z_k}-1\}\{Y_k-m(\mathbf{X}_k,\hat{\beta})\}].
\end{align*}
Under the MAR assumption, we have $Y\perp R|X$. Hence, we have
\begin{align*}
	\hat{\mu}_{AIPWT}&\overset{p}{\rightarrow}\mu+E[\{\frac{R_k}{Z_k}-1\}\{Y_k-m(\mathbf{X}_k,\hat{\beta})\}]\\
	&=\mu+E[E[\{\frac{R_k}{Z_k}-1\}\{Y_k-m(\mathbf{X}_k,\hat{\beta})\}|\mathbf{X}_k]]\\
	&=\mu+E[E[(\frac{R_k}{Z_k}-1)|\mathbf{X}_k]E[(Y_k-m(\mathbf{X}_k,\hat{\beta}))|\mathbf{X}_k]].
\end{align*}

Suppose that the true propensity model is $\pi_0(X)$ and the propensity model in equation (\ref{rrz}) is correctly specified. Then $Z_k\overset{p}{\rightarrow} \pi_0(\mathbf{X}_k)$ and
\begin{align*}
	E[(\frac{R_k}{Z_k}-1)|\mathbf{X}_k]&\overset{p}{\rightarrow} E[(\frac{R_k}{\pi_0(\mathbf{X}_k)}-1)|\mathbf{X}_k]\\
	&=\frac{\pi_0(\mathbf{X}_k)}{\pi_0(\mathbf{X}_k)}-1\\
	&=0.
\end{align*}
This implies that $\hat{\mu}_{AIPWT}\overset{p}{\rightarrow}\mu$ if the propensity model is correctly specified regardless of whether the mean model is correctly specified. Now suppose that the true mean model is $m_0(\mathbf{X}_k)$ and the mean model in equation (\ref{rrz}) is correctly specified. Then $m(\mathbf{X}_k,\hat{\beta})\overset{p}{\rightarrow}m_0(\mathbf{X}_k)$ and
\begin{align*}
	E[(Y_k-m(\mathbf{X}_k,\hat{\beta}))|\mathbf{X}_k]&\overset{p}{\rightarrow} E[(Y_k-m_0(\mathbf{X}_k))|\mathbf{X}_k]\\
	&=\mu-\mu\\
	&=0.
\end{align*}
Hence, $\hat{\mu}_{AIPWT}\overset{p}{\rightarrow}\mu$ if the mean model is correctly specified.

\noindent
{\bf{Web Appendix B: Consistency of the PSPP estimator}}

We show that the PSPP model is doubly robust closely following \cite{zhang_little}'s arguments in the first corollary of their supplementary materials. We first rewrite equation (\ref{pspp}) as
\begin{equation}
	\label{pspp_dr}
	(Y_k|Z_k,X_{k1},\ldots,X_{kp};\mathbf{\phi},\eta)\sim N(s(Z_k;\mathbf{\phi})+f(X_{k1},\ldots,X_{kp},\eta),\sigma^2),
\end{equation}
where $s[Z_k;\mathbf{\phi}]=\phi_0+\sum_{l=1}^L\phi_lZ_k^L+\sum_{h=1}^H\phi_{L+h}(Z_k-\tau_h)_+^L$. Suppose we specified the mean function $f(X_{k1},\ldots,X_{kp},\eta)$ correctly, then $s(Z_k;\mathbf{\phi})$ is absorbed into the error term and hence $s(Z_k;\mathbf{\phi})+f(X_{k1},\ldots,X_{kp},\eta)\overset{p}{\rightarrow}\mu$.

Now suppose instead that equation (\ref{prop_freq}) was specified correctly. We consider two scenarios, one where we omit $f(X_{k1},\ldots,X_{kp},\eta)$ in equation (\ref{pspp}) and the other where $f(X_{k1},\ldots,X_{kp},\eta)$ is specified. Let
\[
	\mathbf{Z}=[1,Z_k,(Z_k-\tau_1)_+,\ldots,(Z_k-\tau_L)_+],
\]
the truncated linear basis of the propensity score and
\[
	\mathbf{X}=[f_1(X_{k1},\ldots,X_{kp}),\ldots,f_T(X_{k1},\ldots,X_{kp})]=[V_{k1},\ldots,V_{kT}]
\]
be the elements in the function $f$. Let $T$ be the total number of elements in $f$. For the scenario where we omit $f(X_{k1},\ldots,X_{kp},\eta)$, $E[Y_k|Z_k]=\phi_0+\phi_1Z_k+\sum_{h=1}^H\phi_{1+h}(Z_k-\tau_h)_+$ and we obtain $\mathbf{\phi}$ by minimizing  $||\mathbf{Y}-\mathbf{Z}\mathbf{\phi}||^2+\lambda^2\mathbf{\phi}^T\mathbf{D}\mathbf{\phi}$ where $\phi=(\phi_0, \phi_1,\ldots,\phi_{1+H})^T$, $\lambda$ is the penalty, and $\mathbf{D}=\text{diag}(1_H)$. Using the mixed model representation and by restricted maximum likelihood estimation, $\hat{\mathbf{Y}}(Z_k,\hat{\lambda},\mathbf{D})=\mathbf{Z}(\mathbf{Z}^T\mathbf{Z}+\hat{\lambda}^2\mathbf{D})^{-1}\mathbf{Z}^T\mathbf{Y}$. As $n\rightarrow\infty$, $\hat{\lambda}\rightarrow 0$ and hence the predicted value of $\mathbf{Y}$ converges to
\begin{equation}
	\label{pspp_pred3}
	\hat{\mathbf{Y}}(\mathbf{Z},0,\mathbf{D})=\mathbf{Z}(\mathbf{Z}^T\mathbf{Z})^{-1}\mathbf{Z}^T\mathbf{Y}=\mathbf{Z}\hat{\mathbf{\phi}}.
\end{equation}
Equation (\ref{pspp_pred3}) estimates the marginal mean of $\mathbf{Y}$ consistently because of the balancing property of propensity score, $Z_k$, that is, missingness is completely at random conditional on $Z_k$, so predicted values of $Y_k$ using a smooth function of $Z$ should yield consistent estimation of the missing values.

If $f(X_{k1},\ldots,X_{kp}),\eta)$ was specified but incorrect, then the conditional mean of $\mathbf{Y}$ is
\begin{align*}
	E[Y_k|Z_k,X_{k1},\ldots,X_{kp}]&=s(Z_k;\mathbf{\phi})+f(X_{k1},\ldots,X_{kp},\mathbf{\eta})\\
	&=\phi_0+\phi_1Z_k+\sum_{h=1}^H\phi_{1+h}(Z_k-\tau_h)_++\mathbf{X}\mathbf{\eta}.
\end{align*}
$(\mathbf{\phi},\mathbf{\eta})^T$ is obtained by minimizing $||\mathbf{Y}-[\mathbf{Z},\mathbf{X}](\mathbf{\phi},\mathbf{\eta})^T||^2+\lambda^2(\mathbf{\phi},\mathbf{\eta})\mathbf{D}(\mathbf{\phi},\mathbf{\eta})^T$ where $\lambda$ is the penalty and $\mathbf{D}=\text{diag}(1_H,0_{2+T})$. Using the mixed model representation and by restricted maximum likelihood estimation, $\hat{\mathbf{Y}}(Z_k,X_{k1},\ldots,X_{kp},\hat{\lambda},\mathbf{D})=[\mathbf{Z},\mathbf{X}]([\mathbf{Z},\mathbf{X}]^T[\mathbf{Z},\mathbf{X}]+\hat{\lambda}^2\mathbf{D})^{-1}[\mathbf{Z},\mathbf{X}]^T\mathbf{Y}$.

When $n\rightarrow\infty$, $\hat{\lambda}\rightarrow 0$ and $\hat{\mathbf{Y}}(Z_k,X_{k1},\ldots,X_{kp},\hat{\lambda},\mathbf{D})\rightarrow [\mathbf{Z},\mathbf{X}]([\mathbf{Z},\mathbf{X}]^T[\mathbf{Z},\mathbf{X}])^{-1}[\mathbf{Z},\mathbf{X}]^T\mathbf{Y}$. The predicted value of $\mathbf{Y}$ can then be written as
\begin{equation}
	\label{pspp_pred1}
	\hat{\mathbf{Y}}(Z_k,X_{k1},\ldots,X_{kp},0,\mathbf{D})=\mathbf{Z}\hat{\mathbf{\phi}}+\mathbf{X}\hat{\mathbf{\eta}}.
\end{equation}
Now we regress each term in $f$ on the propensity score i.e. $V_i$ on $\mathbf{Z}$ for all $i=1,\ldots,T$ where $\mathbf{Z}$ is the predictor and each $V_i$ are the outcome. As $n\rightarrow\infty$, the predicted value of each element in $f$, $\hat{\mathbf{V}}_i(\mathbf{Z};\hat{\lambda})\rightarrow\hat{\mathbf{V}}_i(\mathbf{Z};0)\rightarrow\mathbf{Z}(\mathbf{Z}^T\mathbf{Z})^{-1}\mathbf{Z}\mathbf{V}_i$. Let $\hat{\mathbf{X}}=[\mathbf{V}_1,\ldots,\mathbf{V}_T]$ and substitute $\hat{\mathbf{X}}$ into equation (\ref{pspp_pred1}). Then
\begin{equation}
	\label{pspp_pred2}
	E[\hat{\mathbf{Y}}(Z_k,X_{k1},\ldots,X_{kp})|Z_k]=\mathbf{Z}\hat{\mathbf{\phi}}+\hat{\mathbf{X}}\hat{\mathbf{\eta}}.
\end{equation}
By lemma 1 in \cite{zhang_little}'s supplementary materials, equation (\ref{pspp_pred2}) converges to equation (\ref{pspp_pred3}) as $n\rightarrow\infty$ and hence equation (\ref{pspp}) is consistent for the marginal mean of $Y$ if the propensity model is correctly specified but the mean model is incorrectly specified.

\noindent
{\bf{Web Appendix C: Simulation Results for Sample Sizes 500, 1,000, and 5,000}}

\noindent
{\bf{Linear interaction in mean model}}

\begin{table}
\caption{Bias, RMSE, 95\% coverage, and average 95\% confidence interval length (AIL) of the eight estimators under the linear interaction in mean model scenario with sample size 500 using bootstrap.\label{app_lin_500}}
	\begin{center}
	\hspace*{-1cm}
\begin{tabular}{|lcccc|cccc|}
		\hline
		$n=500$ & \multicolumn{4}{c|}{Both correct} & \multicolumn{4}{c|}{Propensity correct} \\
		Method & Bias & RMSE & Coverage & AIL & Bias & RMSE & Coverage & AIL \\ \hline
		BD	&	-0.01	&	0.13	&	94	&	0.48	&	-0.01	&	0.13	&	94	&	0.48	\\
CC	&	0.51	&	0.53	&	9.8	&	0.59	&	0.51	&	0.53	&	9.8	&	0.59	\\
MLR	&	0	&	0.17	&	98.6	&	0.87	&	0.44	&	0.47	&	40.6	&	0.81	\\
PSPP	&	-0.01	&	0.21	&	99	&	1.29	&	0.05	&	0.18	&	97.4	&	0.86	\\
AIPWT	&	0	&	0.18	&	93.4	&	0.67	&	0.03	&	0.36	&	89.2	&	1.04	\\
PSBPP	&	0	&	0.18	&	98.8	&	0.95	&	-0.06	&	0.21	&	98.6	&	1.1	\\
AIPWT with BART	&	0.15	&	0.23	&	78.4	&	0.61	&	0.15	&	0.23	&	78.4	&	0.61	\\
BART	&	0.19	&	0.26	&	86.8	&	0.79	&	0.19	&	0.26	&	86.8	&	0.79	\\
BARTps	&	0.11	&	0.21	&	93.6	&	0.83	&	0.11	&	0.21	&	93.6	&	0.83	\\
		\hline
		& \multicolumn{4}{c|}{Mean correct} & \multicolumn{4}{c|}{Both wrong}\\
		Method & Bias & RMSE & Coverage & AIL & Bias & RMSE & Coverage & AIL \\
		\hline
		BD	&	-0.01	&	0.13	&	94	&	0.48	&	-0.01	&	0.13	&	94	&	0.48	\\
CC	&	0.51	&	0.53	&	9.8	&	0.59	&	0.51	&	0.53	&	9.8	&	0.59	\\
MLR	&	0	&	0.17	&	98.6	&	0.87	&	0.44	&	0.47	&	40.6	&	0.81	\\
PSPP	&	0	&	0.18	&	98.8	&	0.9	&	0.25	&	0.31	&	90.4	&	1.04	\\
AIPWT	&	0	&	0.18	&	93	&	0.65	&	0.42	&	0.46	&	25.8	&	0.61	\\
PSBPP	&	0	&	0.18	&	98.8	&	0.95	&	-0.06	&	0.21	&	98.6	&	1.1	\\
AIPWT with BART	&	0.15	&	0.23	&	78.4	&	0.61	&	0.15	&	0.23	&	78.4	&	0.61	\\
BART	&	0.19	&	0.26	&	86.8	&	0.79	&	0.19	&	0.26	&	86.8	&	0.79	\\
BARTps	&	0.11	&	0.21	&	93.6	&	0.83	&	0.11	&	0.21	&	93.6	&	0.83	\\
		\hline
	\end{tabular}
\end{center}
\end{table}

\begin{table}
\caption{Bias, RMSE, 95\% coverage, and average 95\% confidence interval length (AIL) of the eight estimators under the linear interaction in mean model scenario with sample size 1,000 using bootstrap.\label{app_lin_1000}}
\begin{center}
	\hspace*{-1cm}
	\begin{tabular}{|lcccc|cccc|}
		\hline
		$n=1,000$ & \multicolumn{4}{c|}{Both correct} & \multicolumn{4}{c|}{Propensity correct} \\
		Method & Bias & RMSE & Coverage & AIL & Bias & RMSE & Coverage & AIL \\ \hline
		BD	&	0	&	0.09	&	94.2	&	0.34	&	0	&	0.09	&	94.2	&	0.34	\\
CC	&	0.51	&	0.53	&	0.6	&	0.42	&	0.51	&	0.53	&	0.6	&	0.42	\\
MLR	&	0	&	0.12	&	99	&	0.62	&	0.45	&	0.46	&	10	&	0.57	\\
PSPP	&	0.01	&	0.14	&	99.8	&	0.78	&	0.05	&	0.13	&	97.4	&	0.61	\\
AIPWT	&	0	&	0.12	&	94.4	&	0.47	&	0.04	&	0.18	&	87.2	&	0.6	\\
PSBPP	&	0	&	0.13	&	99.2	&	0.64	&	-0.06	&	0.15	&	98.4	&	0.71	\\
AIPWT with BART	&	0.11	&	0.17	&	78.2	&	0.44	&	0.11	&	0.17	&	78.2	&	0.44	\\
BART	&	0.14	&	0.19	&	87.4	&	0.57	&	0.14	&	0.19	&	87.4	&	0.57	\\
BARTps	&	0.07	&	0.14	&	95.8	&	0.6	&	0.07	&	0.14	&	95.8	&	0.6	\\
		\hline
		& \multicolumn{4}{c|}{Mean correct} & \multicolumn{4}{c|}{Both wrong}\\
		Method & Bias & RMSE & Coverage & AIL & Bias & RMSE & Coverage & AIL \\
		\hline
		BD	&	0	&	0.09	&	94.2	&	0.34	&	0	&	0.09	&	94.2	&	0.34	\\
CC	&	0.51	&	0.53	&	0.6	&	0.42	&	0.51	&	0.53	&	0.6	&	0.42	\\
MLR	&	0	&	0.12	&	99	&	0.62	&	0.45	&	0.46	&	10	&	0.57	\\
PSPP	&	0	&	0.12	&	99	&	0.63	&	0.22	&	0.26	&	85	&	0.78	\\
AIPWT	&	0	&	0.12	&	92.8	&	0.46	&	0.43	&	0.45	&	5	&	0.43	\\
PSBPP	&	0	&	0.13	&	99.2	&	0.64	&	-0.06	&	0.15	&	98.4	&	0.71	\\
AIPWT with BART	&	0.11	&	0.17	&	78.2	&	0.44	&	0.11	&	0.17	&	78.2	&	0.44	\\
BART	&	0.14	&	0.19	&	87.4	&	0.57	&	0.14	&	0.19	&	87.4	&	0.57	\\
BARTps	&	0.07	&	0.14	&	95.8	&	0.6	&	0.07	&	0.14	&	95.8	&	0.6	\\
		\hline
	\end{tabular}
\end{center}
\end{table}

\begin{table}
\caption{Bias, RMSE, 95\% coverage, and average 95\% confidence interval length (AIL) of the eight estimators under the linear interaction in mean model scenario with sample size 5,000 using bootstrap.\label{app_lin_5000}}
\begin{center}
	\hspace*{-1cm}
	\begin{tabular}{|lcccc|cccc|}
		\hline
		$n=5,000$ & \multicolumn{4}{c|}{Both correct} & \multicolumn{4}{c|}{Propensity correct} \\
		Method & Bias & RMSE & Coverage & AIL & Bias & RMSE & Coverage & AIL \\ \hline
		BD	&	0	&	0.04	&	96	&	0.15	&	0	&	0.04	&	96	&	0.15	\\
CC	&	0.51	&	0.52	&	0	&	0.19	&	0.51	&	0.52	&	0	&	0.19	\\
MLR	&	0	&	0.05	&	99.4	&	0.27	&	0.44	&	0.45	&	0	&	0.26	\\
PSPP	&	0	&	0.05	&	99.2	&	0.29	&	0.03	&	0.06	&	97.6	&	0.27	\\
AIPWT	&	0	&	0.05	&	94.4	&	0.21	&	0.01	&	0.1	&	88.4	&	0.32	\\
PSBPP	&	0	&	0.05	&	99.4	&	0.28	&	-0.04	&	0.07	&	97.2	&	0.29	\\
AIPWT with BART	&	0.05	&	0.07	&	80.6	&	0.2	&	0.05	&	0.07	&	80.6	&	0.2	\\
BART	&	0.06	&	0.08	&	88.8	&	0.27	&	0.06	&	0.08	&	88.8	&	0.27	\\
BARTps	&	0.02	&	0.06	&	97.6	&	0.27	&	0.02	&	0.06	&	97.6	&	0.27	\\
		\hline
		& \multicolumn{4}{c|}{Mean correct} & \multicolumn{4}{c|}{Both wrong}\\
		Method & Bias & RMSE & Coverage & AIL & Bias & RMSE & Coverage & AIL \\
		\hline
		BD	&	0	&	0.04	&	96	&	0.15	&	0	&	0.04	&	96	&	0.15	\\
CC	&	0.51	&	0.52	&	0	&	0.19	&	0.51	&	0.52	&	0	&	0.19	\\
MLR	&	0	&	0.05	&	99.4	&	0.27	&	0.44	&	0.45	&	0	&	0.26	\\
PSPP	&	0	&	0.05	&	99.4	&	0.28	&	0.16	&	0.19	&	69.4	&	0.4	\\
AIPWT	&	0	&	0.05	&	94.8	&	0.2	&	0.43	&	0.43	&	0	&	0.19	\\
PSBPP	&	0	&	0.05	&	99.4	&	0.28	&	-0.04	&	0.07	&	97.2	&	0.29	\\
AIPWT with BART	&	0.05	&	0.07	&	80.6	&	0.2	&	0.05	&	0.07	&	80.6	&	0.2	\\
BART	&	0.06	&	0.08	&	88.8	&	0.27	&	0.06	&	0.08	&	88.8	&	0.27	\\
BARTps	&	0.02	&	0.06	&	97.6	&	0.27	&	0.02	&	0.06	&	97.6	&	0.27	\\
		\hline
	\end{tabular}
\end{center}
\end{table}

\begin{table}
\caption{Bias, RMSE, 95\% coverage, and average 95\% confidence interval length (AIL) of the eight estimators under the linear interaction in mean model scenario with sample size 500 using MI with posterior mean of propensity scores.\label{app_lin_500_mi}}
	\begin{center}
	\hspace*{-1cm}
	\begin{tabular}{|lcccc|cccc|}
		\hline
		$n=500$ & \multicolumn{4}{c|}{Both correct} & \multicolumn{4}{c|}{Propensity correct} \\
		Method & Bias & RMSE & Coverage & AIL & Bias & RMSE & Coverage & AIL \\ \hline
		BD	&	0	&	0.13	&	95	&	0.49	&	0	&	0.13	&	95	&	0.49	\\
CC	&	0.51	&	0.53	&	4	&	0.41	&	0.51	&	0.53	&	4	&	0.41	\\
MLR	&	0	&	0.18	&	94.6	&	0.7	&	0.44	&	0.47	&	27.4	&	0.66	\\
PSPP	&	0	&	0.21	&	96.4	&	0.9	&	0.06	&	0.19	&	92.2	&	0.69	\\
PSBPP	&	0	&	0.18	&	94.6	&	0.76	&	0.01	&	0.2	&	93.8	&	0.79	\\
BART	&	0.18	&	0.26	&	80.6	&	0.68	&	0.18	&	0.26	&	80.6	&	0.68	\\
BARTps	&	0.1	&	0.21	&	90.6	&	0.73	&	0.1	&	0.21	&	90.6	&	0.73	\\
		\hline
		& \multicolumn{4}{c|}{Mean correct} & \multicolumn{4}{c|}{Both wrong}\\
		Method & Bias & RMSE & Coverage & AIL & Bias & RMSE & Coverage & AIL \\
		\hline
		BD	&	0	&	0.13	&	95	&	0.49	&	0	&	0.13	&	95	&	0.49	\\
CC	&	0.51	&	0.53	&	4	&	0.41	&	0.51	&	0.53	&	4	&	0.41	\\
MLR	&	0	&	0.18	&	94.6	&	0.7	&	0.44	&	0.47	&	27.4	&	0.66	\\
PSPP	&	0	&	0.18	&	94.2	&	0.73	&	0.22	&	0.32	&	69.2	&	0.73	\\
PSBPP	&	0	&	0.18	&	94.6	&	0.76	&	0.01	&	0.2	&	93.8	&	0.79	\\
BART	&	0.18	&	0.26	&	80.6	&	0.68	&	0.18	&	0.26	&	80.6	&	0.68	\\
BARTps	&	0.1	&	0.21	&	90.6	&	0.73	&	0.1	&	0.21	&	90.6	&	0.73	\\
		\hline
	\end{tabular}
\end{center}
\end{table}

\begin{table}
\caption{Bias, RMSE, 95\% coverage, and average 95\% confidence interval length (AIL) of the eight estimators under the linear interaction in mean model scenario with sample size 1,000 using MI with posterior mean of propensity scores.\label{app_lin_1000_mi}}
\begin{center}
	\hspace*{-1cm}
	\begin{tabular}{|lcccc|cccc|}
		\hline
		$n=1,000$ & \multicolumn{4}{c|}{Both correct} & \multicolumn{4}{c|}{Propensity correct} \\
		Method & Bias & RMSE & Coverage & AIL & Bias & RMSE & Coverage & AIL \\ \hline
		BD	&	0	&	0.09	&	94.6	&	0.35	&	0	&	0.09	&	94.6	&	0.35	\\
CC	&	0.51	&	0.52	&	0	&	0.29	&	0.51	&	0.52	&	0	&	0.29	\\
MLR	&	0	&	0.12	&	95	&	0.5	&	0.44	&	0.46	&	5.2	&	0.47	\\
PSPP	&	0	&	0.14	&	96.6	&	0.58	&	0.06	&	0.14	&	93.2	&	0.49	\\
PSBPP	&	0	&	0.13	&	95.6	&	0.52	&	0	&	0.14	&	94.4	&	0.55	\\
BART	&	0.14	&	0.19	&	78.6	&	0.49	&	0.14	&	0.19	&	78.6	&	0.49	\\
BARTps	&	0.06	&	0.15	&	90.6	&	0.52	&	0.06	&	0.15	&	90.6	&	0.52	\\
		\hline
		& \multicolumn{4}{c|}{Mean correct} & \multicolumn{4}{c|}{Both wrong}\\
		Method & Bias & RMSE & Coverage & AIL & Bias & RMSE & Coverage & AIL \\
		\hline
		BD	&	0	&	0.09	&	94.6	&	0.35	&	0	&	0.09	&	94.6	&	0.35	\\
CC	&	0.51	&	0.52	&	0	&	0.29	&	0.51	&	0.52	&	0	&	0.29	\\
MLR	&	0	&	0.12	&	95	&	0.5	&	0.44	&	0.46	&	5.2	&	0.47	\\
PSPP	&	0	&	0.13	&	95.2	&	0.51	&	0.2	&	0.27	&	63.8	&	0.51	\\
PSBPP	&	0	&	0.13	&	95.6	&	0.52	&	0	&	0.14	&	94.4	&	0.55	\\
BART	&	0.14	&	0.19	&	78.6	&	0.49	&	0.14	&	0.19	&	78.6	&	0.49	\\
BARTps	&	0.06	&	0.15	&	90.6	&	0.52	&	0.06	&	0.15	&	90.6	&	0.52	\\
		\hline
	\end{tabular}
\end{center}
\end{table}

\begin{table}
\caption{Bias, RMSE, 95\% coverage, and average 95\% confidence interval length (AIL) of the eight estimators under the linear interaction in mean model scenario with sample size 5,000 using MI with posterior mean of propensity scores.\label{app_lin_5000_mi}}
\begin{center}
	\hspace*{-1cm}
	\begin{tabular}{|lcccc|cccc|}
		\hline
		$n=5,000$ & \multicolumn{4}{c|}{Both correct} & \multicolumn{4}{c|}{Propensity correct} \\
		Method & Bias & RMSE & Coverage & AIL & Bias & RMSE & Coverage & AIL \\ \hline
		BD	&	0	&	0.04	&	97	&	0.15	&	0	&	0.04	&	97	&	0.15	\\
CC	&	0.51	&	0.52	&	0	&	0.13	&	0.51	&	0.52	&	0	&	0.13	\\
MLR	&	0	&	0.05	&	96.8	&	0.22	&	0.44	&	0.45	&	0	&	0.21	\\
PSPP	&	0	&	0.05	&	96.4	&	0.24	&	0.03	&	0.07	&	90	&	0.22	\\
PSBPP	&	0	&	0.05	&	96.6	&	0.22	&	0	&	0.06	&	95.4	&	0.23	\\
BART	&	0.06	&	0.08	&	80.2	&	0.22	&	0.06	&	0.08	&	80.2	&	0.22	\\
BARTps	&	0.03	&	0.06	&	94.8	&	0.23	&	0.03	&	0.06	&	94.8	&	0.23	\\
		\hline
		& \multicolumn{4}{c|}{Mean correct} & \multicolumn{4}{c|}{Both wrong}\\
		Method & Bias & RMSE & Coverage & AIL & Bias & RMSE & Coverage & AIL \\
		\hline
		BD	&	0	&	0.04	&	97	&	0.15	&	0	&	0.04	&	97	&	0.15	\\
CC	&	0.51	&	0.52	&	0	&	0.13	&	0.51	&	0.52	&	0	&	0.13	\\
MLR	&	0	&	0.05	&	96.8	&	0.22	&	0.44	&	0.45	&	0	&	0.21	\\
PSPP	&	0	&	0.05	&	96	&	0.22	&	0.16	&	0.19	&	33.6	&	0.22	\\
PSBPP	&	0	&	0.05	&	96.6	&	0.22	&	0	&	0.06	&	95.4	&	0.23	\\
BART	&	0.06	&	0.08	&	80.2	&	0.22	&	0.06	&	0.08	&	80.2	&	0.22	\\
BARTps	&	0.03	&	0.06	&	94.8	&	0.23	&	0.03	&	0.06	&	94.8	&	0.23	\\
		\hline
	\end{tabular}
\end{center}
\end{table}

\begin{table}
\caption{Bias, RMSE, 95\% coverage, and average 95\% confidence interval length (AIL) of the eight estimators under the linear interaction in mean model scenario with sample size 500 using MI with posterior draw of propensity scores.\label{app_lin_500_mi_new}}
	\begin{center}
	\hspace*{-1cm}
	\begin{tabular}{|lcccc|cccc|}
		\hline
		$n=500$ & \multicolumn{4}{c|}{Both correct} & \multicolumn{4}{c|}{Propensity correct} \\
		Method & Bias & RMSE & Coverage & AIL & Bias & RMSE & Coverage & AIL \\ \hline
		BD	&	0	&	0.13	&	95.2	&	0.49	&	0	&	0.13	&	95.2	&	0.49	\\
CC	&	0.51	&	0.53	&	4	&	0.41	&	0.51	&	0.53	&	4	&	0.41	\\
MLR	&	0	&	0.18	&	94.6	&	0.7	&	0.44	&	0.47	&	27.6	&	0.66	\\
PSPP	&	0.01	&	0.2	&	96.6	&	0.92	&	0.06	&	0.18	&	93.6	&	0.69	\\
PSBPP	&	0	&	0.18	&	94.6	&	0.75	&	0.2	&	0.27	&	86.2	&	0.84	\\
BART	&	0.18	&	0.26	&	80.8	&	0.68	&	0.18	&	0.26	&	80.8	&	0.68	\\
BARTps	&	0.17	&	0.25	&	84.2	&	0.71	&	0.17	&	0.25	&	84.2	&	0.71	\\
		\hline
		& \multicolumn{4}{c|}{Mean correct} & \multicolumn{4}{c|}{Both wrong}\\
		Method & Bias & RMSE & Coverage & AIL & Bias & RMSE & Coverage & AIL \\
		\hline
		BD	&	0	&	0.13	&	95.2	&	0.49	&	0	&	0.13	&	95.2	&	0.49	\\
CC	&	0.51	&	0.53	&	4	&	0.41	&	0.51	&	0.53	&	4	&	0.41	\\
MLR	&	0	&	0.18	&	94.6	&	0.7	&	0.44	&	0.47	&	27.6	&	0.66	\\
PSPP	&	0	&	0.18	&	94.2	&	0.73	&	0.25	&	0.32	&	80.6	&	0.91	\\
PSBPP	&	0	&	0.18	&	94.6	&	0.75	&	0.2	&	0.27	&	86.2	&	0.84	\\
BART	&	0.18	&	0.26	&	80.8	&	0.68	&	0.18	&	0.26	&	80.8	&	0.68	\\
BARTps	&	0.17	&	0.25	&	84.2	&	0.71	&	0.17	&	0.25	&	84.2	&	0.71	\\
		\hline
	\end{tabular}
\end{center}
\end{table}

\begin{table}
\caption{Bias, RMSE, 95\% coverage, and average 95\% confidence interval length (AIL) of the eight estimators under the linear interaction in mean model scenario with sample size 1,000 using MI with posterior draw of propensity scores.\label{app_lin_1000_mi_new}}
\begin{center}
	\hspace*{-1cm}
	\begin{tabular}{|lcccc|cccc|}
		\hline
		$n=1,000$ & \multicolumn{4}{c|}{Both correct} & \multicolumn{4}{c|}{Propensity correct} \\
		Method & Bias & RMSE & Coverage & AIL & Bias & RMSE & Coverage & AIL \\ \hline
		BD	&	0	&	0.09	&	94.6	&	0.35	&	0	&	0.09	&	94.6	&	0.35	\\
CC	&	0.51	&	0.53	&	0	&	0.29	&	0.51	&	0.53	&	0	&	0.29	\\
MLR	&	0	&	0.12	&	95	&	0.5	&	0.44	&	0.46	&	5.2	&	0.47	\\
PSPP	&	0	&	0.14	&	96.4	&	0.58	&	0.06	&	0.14	&	91.6	&	0.49	\\
PSBPP	&	0	&	0.13	&	95.4	&	0.52	&	0.16	&	0.2	&	83.8	&	0.6	\\
BART	&	0.14	&	0.19	&	78.4	&	0.49	&	0.14	&	0.19	&	78.4	&	0.49	\\
BARTps	&	0.12	&	0.18	&	84	&	0.51	&	0.12	&	0.18	&	84	&	0.51	\\
		\hline
		& \multicolumn{4}{c|}{Mean correct} & \multicolumn{4}{c|}{Both wrong}\\
		Method & Bias & RMSE & Coverage & AIL & Bias & RMSE & Coverage & AIL \\
		\hline
		BD	&	0	&	0.09	&	94.6	&	0.35	&	0	&	0.09	&	94.6	&	0.35	\\
CC	&	0.51	&	0.53	&	0	&	0.29	&	0.51	&	0.53	&	0	&	0.29	\\
MLR	&	0	&	0.12	&	95	&	0.5	&	0.44	&	0.46	&	5.2	&	0.47	\\
PSPP	&	0	&	0.13	&	95.6	&	0.51	&	0.23	&	0.27	&	79.6	&	0.72	\\
PSBPP	&	0	&	0.13	&	95.4	&	0.52	&	0.16	&	0.2	&	83.8	&	0.6	\\
BART	&	0.14	&	0.19	&	78.4	&	0.49	&	0.14	&	0.19	&	78.4	&	0.49	\\
BARTps	&	0.12	&	0.18	&	84	&	0.51	&	0.12	&	0.18	&	84	&	0.51	\\
		\hline
	\end{tabular}
\end{center}
\end{table}

\begin{table}
\caption{Bias, RMSE, 95\% coverage, and average 95\% confidence interval length (AIL) of the eight estimators under the linear interaction in mean model scenario with sample size 5,000 using MI with posterior draw of propensity scores.\label{app_lin_5000_mi_new}}
\begin{center}
	\hspace*{-1cm}
	\begin{tabular}{|lcccc|cccc|}
		\hline
		$n=5,000$ & \multicolumn{4}{c|}{Both correct} & \multicolumn{4}{c|}{Propensity correct} \\
		Method & Bias & RMSE & Coverage & AIL & Bias & RMSE & Coverage & AIL \\ \hline
		BD	&	0	&	0.04	&	97	&	0.15	&	0	&	0.04	&	97	&	0.15	\\
CC	&	0.51	&	0.52	&	0	&	0.13	&	0.51	&	0.52	&	0	&	0.13	\\
MLR	&	0	&	0.05	&	96.8	&	0.22	&	0.44	&	0.45	&	0	&	0.21	\\
PSPP	&	0	&	0.05	&	97.8	&	0.23	&	0.04	&	0.06	&	92	&	0.23	\\
PSBPP	&	0	&	0.05	&	95.4	&	0.23	&	0.08	&	0.09	&	82	&	0.26	\\
BART	&	0.06	&	0.08	&	80.2	&	0.22	&	0.06	&	0.08	&	80.2	&	0.22	\\
BARTps	&	0.05	&	0.07	&	87.2	&	0.23	&	0.05	&	0.07	&	87.2	&	0.23	\\
		\hline
		& \multicolumn{4}{c|}{Mean correct} & \multicolumn{4}{c|}{Both wrong}\\
		Method & Bias & RMSE & Coverage & AIL & Bias & RMSE & Coverage & AIL \\
		\hline
		BD	&	0	&	0.04	&	97	&	0.15	&	0	&	0.04	&	97	&	0.15	\\
CC	&	0.51	&	0.52	&	0	&	0.13	&	0.51	&	0.52	&	0	&	0.13	\\
MLR	&	0	&	0.05	&	96.8	&	0.22	&	0.44	&	0.45	&	0	&	0.21	\\
PSPP	&	0	&	0.05	&	97.4	&	0.22	&	0.17	&	0.19	&	63.6	&	0.38	\\
PSBPP	&	0	&	0.05	&	95.4	&	0.23	&	0.08	&	0.09	&	82	&	0.26	\\
BART	&	0.06	&	0.08	&	80.2	&	0.22	&	0.06	&	0.08	&	80.2	&	0.22	\\
BARTps	&	0.05	&	0.07	&	87.2	&	0.23	&	0.05	&	0.07	&	87.2	&	0.23	\\
		\hline
	\end{tabular}
\end{center}
\end{table}

\noindent
{\bf{Quadratic interaction in mean model}}

\begin{table}
\caption{Bias, RMSE, 95\% coverage, and average 95\% confidence interval length (AIL) of the eight estimators under the quadratic interaction in mean model scenario with sample size 500 using bootstrap.\label{app_quad_500}}
\begin{center}
	\hspace*{-1cm}
	\begin{tabular}{|lcccc|cccc|}
		\hline
		$n=500$ & \multicolumn{4}{c|}{Both correct} & \multicolumn{4}{c|}{Propensity correct} \\
		Method & Bias & RMSE & Coverage & AIL & Bias & RMSE & Coverage & AIL \\ \hline
		BD	&	0.02	&	0.34	&	90.4	&	1.19	&	0.02	&	0.34	&	90.4	&	1.19	\\
CC	&	1.21	&	1.23	&	2.8	&	0.89	&	1.21	&	1.23	&	2.8	&	0.89	\\
MLR	&	0.01	&	0.36	&	96.8	&	1.84	&	1.24	&	1.26	&	3.4	&	1.15	\\
PSPP	&	0.01	&	0.37	&	97.2	&	1.87	&	0.31	&	0.54	&	82.6	&	2.39	\\
AIPWT	&	0.02	&	0.36	&	91.2	&	1.31	&	0.16	&	1.75	&	63.8	&	2.75	\\
PSBPP	&	0.01	&	0.37	&	97.4	&	1.87	&	0.21	&	0.49	&	92.4	&	3.03	\\
AIPWT with BART	&	0.57	&	0.67	&	35	&	1.02	&	0.57	&	0.67	&	35	&	1.02	\\
BART	&	0.64	&	0.71	&	46.8	&	1.28	&	0.64	&	0.71	&	46.8	&	1.28	\\
BARTps	&	0.54	&	0.63	&	60.4	&	1.41	&	0.54	&	0.63	&	60.4	&	1.41	\\
		\hline
		& \multicolumn{4}{c|}{Mean correct} & \multicolumn{4}{c|}{Both wrong}\\
		Method & Bias & RMSE & Coverage & AIL & Bias & RMSE & Coverage & AIL \\
		\hline
		BD	&	0.02	&	0.34	&	90.4	&	1.19	&	0.02	&	0.34	&	90.4	&	1.19	\\
CC	&	1.21	&	1.23	&	2.8	&	0.89	&	1.21	&	1.23	&	2.8	&	0.89	\\
MLR	&	0.01	&	0.36	&	96.8	&	1.84	&	1.24	&	1.26	&	3.4	&	1.15	\\
PSPP	&	0.01	&	0.36	&	97	&	1.87	&	0.83	&	0.89	&	61.4	&	2.05	\\
AIPWT	&	0.02	&	0.36	&	91.4	&	1.3	&	1.21	&	1.23	&	2	&	0.84	\\
PSBPP	&	0.01	&	0.37	&	97.4	&	1.87	&	0.21	&	0.49	&	92.4	&	3.03	\\
AIPWT with BART	&	0.57	&	0.67	&	35	&	1.02	&	0.57	&	0.67	&	35	&	1.02	\\
BART	&	0.64	&	0.71	&	46.8	&	1.28	&	0.64	&	0.71	&	46.8	&	1.28	\\
BARTps	&	0.54	&	0.63	&	60.4	&	1.41	&	0.54	&	0.63	&	60.4	&	1.41	\\
		\hline
	\end{tabular}
\end{center}
\end{table}

\begin{table}
\caption{Bias, RMSE, 95\% coverage, and average 95\% confidence interval length (AIL) of the eight estimators under the quadratic interaction in mean model scenario with sample size 1,000 using bootstrap.\label{app_quad_1000}}
\begin{center}
	\hspace*{-1cm}
	\begin{tabular}{|lcccc|cccc|}
		\hline
		$n=1,000$ & \multicolumn{4}{c|}{Both correct} & \multicolumn{4}{c|}{Propensity correct} \\
		Method & Bias & RMSE & Coverage & AIL & Bias & RMSE & Coverage & AIL \\ \hline
		BD	&	0	&	0.24	&	91.8	&	0.86	&	0	&	0.24	&	91.8	&	0.86	\\
CC	&	1.21	&	1.23	&	0.2	&	0.63	&	1.21	&	1.23	&	0.2	&	0.63	\\
MLR	&	0	&	0.26	&	99	&	1.32	&	1.24	&	1.25	&	0	&	0.8	\\
PSPP	&	0	&	0.26	&	98.8	&	1.33	&	0.21	&	0.44	&	81.2	&	2	\\
AIPWT	&	0	&	0.26	&	91.2	&	0.93	&	0.22	&	0.72	&	67	&	1.68	\\
PSBPP	&	0	&	0.26	&	98.6	&	1.33	&	0.13	&	0.35	&	94	&	2.16	\\
AIPWT with BART	&	0.45	&	0.51	&	29.8	&	0.77	&	0.45	&	0.51	&	29.8	&	0.77	\\
BART	&	0.52	&	0.57	&	42	&	0.97	&	0.52	&	0.57	&	42	&	0.97	\\
BARTps	&	0.41	&	0.47	&	63.4	&	1.07	&	0.41	&	0.47	&	63.4	&	1.07	\\
		\hline
		& \multicolumn{4}{c|}{Mean correct} & \multicolumn{4}{c|}{Both wrong}\\
		Method & Bias & RMSE & Coverage & AIL & Bias & RMSE & Coverage & AIL \\
		\hline
		BD	&	0	&	0.24	&	91.8	&	0.86	&	0	&	0.24	&	91.8	&	0.86	\\
CC	&	1.21	&	1.23	&	0.2	&	0.63	&	1.21	&	1.23	&	0.2	&	0.63	\\
MLR	&	0	&	0.26	&	99	&	1.32	&	1.24	&	1.25	&	0	&	0.8	\\
PSPP	&	0	&	0.26	&	98.6	&	1.33	&	0.72	&	0.77	&	61.8	&	1.69	\\
AIPWT	&	0	&	0.25	&	91	&	0.92	&	1.21	&	1.22	&	0	&	0.59	\\
PSBPP	&	0	&	0.26	&	98.6	&	1.33	&	0.13	&	0.35	&	94	&	2.16	\\
AIPWT with BART	&	0.45	&	0.51	&	29.8	&	0.77	&	0.45	&	0.51	&	29.8	&	0.77	\\
BART	&	0.52	&	0.57	&	42	&	0.97	&	0.52	&	0.57	&	42	&	0.97	\\
BARTps	&	0.41	&	0.47	&	63.4	&	1.07	&	0.41	&	0.47	&	63.4	&	1.07	\\
		\hline
	\end{tabular}
\end{center}
\end{table}

\begin{table}
\caption{Bias, RMSE, 95\% coverage, and average 95\% confidence interval length (AIL) of the eight estimators under the quadratic interaction in mean model scenario with sample size 5,000 using bootstrap.\label{app_quad_5000}}
\begin{center}
	\hspace*{-1cm}
	\begin{tabular}{|lcccc|cccc|}
		\hline
		$n=5,000$ & \multicolumn{4}{c|}{Both correct} & \multicolumn{4}{c|}{Propensity correct} \\
		Method & Bias & RMSE & Coverage & AIL & Bias & RMSE & Coverage & AIL \\ \hline
		BD	&	0.01	&	0.1	&	94	&	0.39	&	0.01	&	0.1	&	94	&	0.39	\\
CC	&	1.21	&	1.21	&	0	&	0.29	&	1.21	&	1.21	&	0	&	0.29	\\
MLR	&	0.01	&	0.11	&	98.6	&	0.59	&	1.24	&	1.24	&	0	&	0.37	\\
PSPP	&	0.01	&	0.11	&	98.6	&	0.59	&	0.12	&	0.23	&	80.8	&	0.87	\\
AIPWT	&	0.01	&	0.11	&	95	&	0.42	&	0.09	&	0.45	&	71.8	&	1.16	\\
PSBPP	&	0.01	&	0.11	&	98.6	&	0.59	&	0.09	&	0.17	&	91.2	&	0.9	\\
AIPWT with BART	&	0.24	&	0.26	&	26	&	0.39	&	0.24	&	0.26	&	26	&	0.39	\\
BART	&	0.28	&	0.3	&	40.6	&	0.5	&	0.28	&	0.3	&	40.6	&	0.5	\\
BARTps	&	0.2	&	0.23	&	67.8	&	0.54	&	0.2	&	0.23	&	67.8	&	0.54	\\
		\hline
		& \multicolumn{4}{c|}{Mean correct} & \multicolumn{4}{c|}{Both wrong}\\
		Method & Bias & RMSE & Coverage & AIL & Bias & RMSE & Coverage & AIL \\
		\hline
		BD	&	0.01	&	0.1	&	94	&	0.39	&	0.01	&	0.1	&	94	&	0.39	\\
CC	&	1.21	&	1.21	&	0	&	0.29	&	1.21	&	1.21	&	0	&	0.29	\\
MLR	&	0.01	&	0.11	&	98.6	&	0.59	&	1.24	&	1.24	&	0	&	0.37	\\
PSPP	&	0.01	&	0.11	&	98.6	&	0.59	&	0.52	&	0.56	&	50.6	&	0.99	\\
AIPWT	&	0.01	&	0.11	&	93.8	&	0.42	&	1.21	&	1.21	&	0	&	0.27	\\
PSBPP	&	0.01	&	0.11	&	98.6	&	0.59	&	0.09	&	0.17	&	91.2	&	0.9	\\
AIPWT with BART	&	0.24	&	0.26	&	26	&	0.39	&	0.24	&	0.26	&	26	&	0.39	\\
BART	&	0.28	&	0.3	&	40.6	&	0.5	&	0.28	&	0.3	&	40.6	&	0.5	\\
BARTps	&	0.2	&	0.23	&	67.8	&	0.54	&	0.2	&	0.23	&	67.8	&	0.54	\\
		\hline
	\end{tabular}
\end{center}
\end{table}

\begin{table}
\caption{Bias, RMSE, 95\% coverage, and average 95\% confidence interval length (AIL) of the eight estimators under the quadratic interaction in mean model scenario with sample size 500 using MI with posterior mean of propensity scores.\label{app_quad_500_mi}}
\begin{center}
	\hspace*{-1cm}
	\begin{tabular}{|lcccc|cccc|}
		\hline
		$n=500$ & \multicolumn{4}{c|}{Both correct} & \multicolumn{4}{c|}{Propensity correct} \\
		Method & Bias & RMSE & Coverage & AIL & Bias & RMSE & Coverage & AIL \\ \hline
		BD	&	0.02	&	0.34	&	91	&	1.23	&	0.02	&	0.34	&	91	&	1.23	\\
CC	&	1.2	&	1.23	&	0.8	&	0.62	&	1.2	&	1.23	&	0.8	&	0.62	\\
MLR	&	0.01	&	0.37	&	93.2	&	1.37	&	1.24	&	1.26	&	2	&	1	\\
PSPP	&	0.01	&	0.36	&	93	&	1.39	&	0.28	&	0.65	&	61.2	&	1.18	\\
PSBPP	&	0.01	&	0.37	&	93.8	&	1.4	&	0.25	&	0.66	&	68.4	&	1.31	\\
BART	&	0.59	&	0.68	&	35	&	0.94	&	0.59	&	0.68	&	35	&	0.94	\\
BARTps	&	0.47	&	0.59	&	49.8	&	1.04	&	0.47	&	0.59	&	49.8	&	1.04	\\
		\hline
		& \multicolumn{4}{c|}{Mean correct} & \multicolumn{4}{c|}{Both wrong}\\
		Method & Bias & RMSE & Coverage & AIL & Bias & RMSE & Coverage & AIL \\
		\hline
		BD	&	0.02	&	0.34	&	91	&	1.23	&	0.02	&	0.34	&	91	&	1.23	\\
CC	&	1.2	&	1.23	&	0.8	&	0.62	&	1.2	&	1.23	&	0.8	&	0.62	\\
MLR	&	0.01	&	0.37	&	93.2	&	1.37	&	1.24	&	1.26	&	2	&	1	\\
PSPP	&	0.02	&	0.37	&	93.2	&	1.39	&	0.75	&	0.87	&	36.4	&	1.16	\\
PSBPP	&	0.01	&	0.37	&	93.8	&	1.4	&	0.25	&	0.66	&	68.4	&	1.31	\\
BART	&	0.59	&	0.68	&	35	&	0.94	&	0.59	&	0.68	&	35	&	0.94	\\
BARTps	&	0.47	&	0.59	&	49.8	&	1.04	&	0.47	&	0.59	&	49.8	&	1.04	\\
		\hline
	\end{tabular}
\end{center}
\end{table}

\begin{table}
\caption{Bias, RMSE, 95\% coverage, and average 95\% confidence interval length (AIL) of the eight estimators under the quadratic interaction in mean model scenario with sample size 1,000 using MI with posterior mean of propensity scores.\label{app_quad_1000_mi}}
\begin{center}
	\hspace*{-1cm}
	\begin{tabular}{|lcccc|cccc|}
		\hline
		$n=1,000$ & \multicolumn{4}{c|}{Both correct} & \multicolumn{4}{c|}{Propensity correct} \\
		Method & Bias & RMSE & Coverage & AIL & Bias & RMSE & Coverage & AIL \\ \hline
		BD	&	0	&	0.24	&	92.6	&	0.89	&	0	&	0.24	&	92.6	&	0.89	\\
CC	&	1.21	&	1.23	&	0	&	0.44	&	1.21	&	1.23	&	0	&	0.44	\\
MLR	&	0	&	0.26	&	94.2	&	0.98	&	1.24	&	1.25	&	0	&	0.7	\\
PSPP	&	0	&	0.26	&	94.4	&	0.99	&	0.16	&	0.56	&	54.4	&	0.88	\\
PSBPP	&	0	&	0.26	&	93.8	&	0.99	&	0.21	&	0.41	&	68.2	&	0.91	\\
BART	&	0.47	&	0.53	&	31.4	&	0.71	&	0.47	&	0.53	&	31.4	&	0.71	\\
BARTps	&	0.35	&	0.44	&	51	&	0.77	&	0.35	&	0.44	&	51	&	0.77	\\
		\hline
		& \multicolumn{4}{c|}{Mean correct} & \multicolumn{4}{c|}{Both wrong}\\
		Method & Bias & RMSE & Coverage & AIL & Bias & RMSE & Coverage & AIL \\
		\hline
		BD	&	0	&	0.24	&	92.6	&	0.89	&	0	&	0.24	&	92.6	&	0.89	\\
CC	&	1.21	&	1.23	&	0	&	0.44	&	1.21	&	1.23	&	0	&	0.44	\\
MLR	&	0	&	0.26	&	94.2	&	0.98	&	1.24	&	1.25	&	0	&	0.7	\\
PSPP	&	0	&	0.26	&	94	&	0.99	&	0.66	&	0.75	&	28.4	&	0.84	\\
PSBPP	&	0	&	0.26	&	93.8	&	0.99	&	0.21	&	0.41	&	68.2	&	0.91	\\
BART	&	0.47	&	0.53	&	31.4	&	0.71	&	0.47	&	0.53	&	31.4	&	0.71	\\
BARTps	&	0.35	&	0.44	&	51	&	0.77	&	0.35	&	0.44	&	51	&	0.77	\\
		\hline
	\end{tabular}
\end{center}
\end{table}

\begin{table}
\caption{Bias, RMSE, 95\% coverage, and average 95\% confidence interval length (AIL) of the eight estimators under the quadratic interaction in mean model scenario with sample size 5,000 using MI with posterior mean of propensity scores.\label{app_quad_5000_mi}}
\begin{center}
	\hspace*{-1cm}
	\begin{tabular}{|lcccc|cccc|}
		\hline
		$n=5,000$ & \multicolumn{4}{c|}{Both correct} & \multicolumn{4}{c|}{Propensity correct} \\
		Method & Bias & RMSE & Coverage & AIL & Bias & RMSE & Coverage & AIL \\ \hline
		BD	&	0.01	&	0.1	&	94.8	&	0.4	&	0.01	&	0.1	&	94.8	&	0.4	\\
CC	&	1.21	&	1.21	&	0	&	0.2	&	1.21	&	1.21	&	0	&	0.2	\\
MLR	&	0.01	&	0.11	&	95	&	0.44	&	1.24	&	1.24	&	0	&	0.32	\\
PSPP	&	0.01	&	0.11	&	95	&	0.44	&	0.1	&	0.24	&	52	&	0.4	\\
PSBPP	&	0.01	&	0.11	&	95.2	&	0.44	&	0.15	&	0.22	&	56.4	&	0.39	\\
BART	&	0.25	&	0.27	&	29.8	&	0.36	&	0.25	&	0.27	&	29.8	&	0.36	\\
BARTps	&	0.17	&	0.21	&	56.8	&	0.39	&	0.17	&	0.21	&	56.8	&	0.39	\\
		\hline
		& \multicolumn{4}{c|}{Mean correct} & \multicolumn{4}{c|}{Both wrong}\\
		Method & Bias & RMSE & Coverage & AIL & Bias & RMSE & Coverage & AIL \\
		\hline
		BD	&	0.01	&	0.1	&	94.8	&	0.4	&	0.01	&	0.1	&	94.8	&	0.4	\\
CC	&	1.21	&	1.21	&	0	&	0.2	&	1.21	&	1.21	&	0	&	0.2	\\
MLR	&	0.01	&	0.11	&	95	&	0.44	&	1.24	&	1.24	&	0	&	0.32	\\
PSPP	&	0.01	&	0.11	&	95.4	&	0.44	&	0.5	&	0.56	&	15	&	0.39	\\
PSBPP	&	0.01	&	0.11	&	95.2	&	0.44	&	0.15	&	0.22	&	56.4	&	0.39	\\
BART	&	0.25	&	0.27	&	29.8	&	0.36	&	0.25	&	0.27	&	29.8	&	0.36	\\
BARTps	&	0.17	&	0.21	&	56.8	&	0.39	&	0.17	&	0.21	&	56.8	&	0.39	\\
		\hline
	\end{tabular}
\end{center}
\end{table}

\begin{table}
\caption{Bias, RMSE, 95\% coverage, and average 95\% confidence interval length (AIL) of the eight estimators under the quadratic interaction in mean model scenario with sample size 500 using MI with posterior draw of propensity scores.\label{app_quad_500_mi_new}}
\begin{center}
	\hspace*{-1cm}
	\begin{tabular}{|lcccc|cccc|}
		\hline
		$n=500$ & \multicolumn{4}{c|}{Both correct} & \multicolumn{4}{c|}{Propensity correct} \\
		Method & Bias & RMSE & Coverage & AIL & Bias & RMSE & Coverage & AIL \\ \hline
		BD	&	0.01	&	0.34	&	90.8	&	1.24	&	0.01	&	0.34	&	90.8	&	1.24	\\
CC	&	1.2	&	1.23	&	0.8	&	0.62	&	1.2	&	1.23	&	0.8	&	0.62	\\
MLR	&	0.01	&	0.37	&	93	&	1.38	&	1.23	&	1.26	&	2	&	1	\\
PSPP	&	0.01	&	0.37	&	92.4	&	1.39	&	0.3	&	0.52	&	72	&	1.46	\\
PSBPP	&	0.01	&	0.37	&	93.6	&	1.41	&	0.67	&	0.79	&	54	&	2.18	\\
BART	&	0.59	&	0.68	&	35.4	&	0.94	&	0.59	&	0.68	&	35.4	&	0.94	\\
BARTps	&	0.57	&	0.66	&	40.4	&	1.03	&	0.57	&	0.66	&	40.4	&	1.03	\\
		\hline
		& \multicolumn{4}{c|}{Mean correct} & \multicolumn{4}{c|}{Both wrong}\\
		Method & Bias & RMSE & Coverage & AIL & Bias & RMSE & Coverage & AIL \\
		\hline
		BD	&	0.01	&	0.34	&	90.8	&	1.24	&	0.01	&	0.34	&	90.8	&	1.24	\\
CC	&	1.2	&	1.23	&	0.8	&	0.62	&	1.2	&	1.23	&	0.8	&	0.62	\\
MLR	&	0.01	&	0.37	&	93	&	1.38	&	1.23	&	1.26	&	2	&	1	\\
PSPP	&	0.01	&	0.37	&	93.6	&	1.41	&	0.79	&	0.86	&	50.4	&	1.72	\\
PSBPP	&	0.01	&	0.37	&	93.6	&	1.41	&	0.67	&	0.79	&	54	&	2.18	\\
BART	&	0.59	&	0.68	&	35.4	&	0.94	&	0.59	&	0.68	&	35.4	&	0.94	\\
BARTps	&	0.57	&	0.66	&	40.4	&	1.03	&	0.57	&	0.66	&	40.4	&	1.03	\\
		\hline
	\end{tabular}
\end{center}
\end{table}

\begin{table}
\caption{Bias, RMSE, 95\% coverage, and average 95\% confidence interval length (AIL) of the eight estimators under the quadratic interaction in mean model scenario with sample size 1,000 using MI with posterior draw of propensity scores.\label{app_quad_1000_mi_new}}
\begin{center}
	\hspace*{-1cm}
	\begin{tabular}{|lcccc|cccc|}
		\hline
		$n=1,000$ & \multicolumn{4}{c|}{Both correct} & \multicolumn{4}{c|}{Propensity correct} \\
		Method & Bias & RMSE & Coverage & AIL & Bias & RMSE & Coverage & AIL \\ \hline
		BD	&	0	&	0.24	&	92.2	&	0.89	&	0	&	0.24	&	92.2	&	0.89	\\
CC	&	1.21	&	1.23	&	0	&	0.44	&	1.21	&	1.23	&	0	&	0.44	\\
MLR	&	0	&	0.26	&	94	&	0.98	&	1.24	&	1.25	&	0	&	0.7	\\
PSPP	&	0	&	0.26	&	93.6	&	0.99	&	0.19	&	0.45	&	71.4	&	1.24	\\
PSBPP	&	0	&	0.26	&	93	&	0.99	&	0.53	&	0.64	&	58.6	&	2.07	\\
BART	&	0.47	&	0.53	&	30.6	&	0.71	&	0.47	&	0.53	&	30.6	&	0.71	\\
BARTps	&	0.44	&	0.51	&	39	&	0.78	&	0.44	&	0.51	&	39	&	0.78	\\
		\hline
		& \multicolumn{4}{c|}{Mean correct} & \multicolumn{4}{c|}{Both wrong}\\
		Method & Bias & RMSE & Coverage & AIL & Bias & RMSE & Coverage & AIL \\
		\hline
		BD	&	0	&	0.24	&	92.2	&	0.89	&	0	&	0.24	&	92.2	&	0.89	\\
CC	&	1.21	&	1.23	&	0	&	0.44	&	1.21	&	1.23	&	0	&	0.44	\\
MLR	&	0	&	0.26	&	94	&	0.98	&	1.24	&	1.25	&	0	&	0.7	\\
PSPP	&	0	&	0.26	&	93.6	&	0.99	&	0.71	&	0.77	&	52.8	&	1.53	\\
PSBPP	&	0	&	0.26	&	93	&	0.99	&	0.53	&	0.64	&	58.6	&	2.07	\\
BART	&	0.47	&	0.53	&	30.6	&	0.71	&	0.47	&	0.53	&	30.6	&	0.71	\\
BARTps	&	0.44	&	0.51	&	39	&	0.78	&	0.44	&	0.51	&	39	&	0.78	\\
		\hline
	\end{tabular}
\end{center}
\end{table}

\begin{table}
\caption{Bias, RMSE, 95\% coverage, and average 95\% confidence interval length (AIL) of the eight estimators under the quadratic interaction in mean model scenario with sample size 5,000 using MI with posterior draw of propensity scores.\label{app_quad_5000_mi_new}}
\begin{center}
	\hspace*{-1cm}
	\begin{tabular}{|lcccc|cccc|}
		\hline
		$n=5,000$ & \multicolumn{4}{c|}{Both correct} & \multicolumn{4}{c|}{Propensity correct} \\
		Method & Bias & RMSE & Coverage & AIL & Bias & RMSE & Coverage & AIL \\ \hline
		BD	&	0.01	&	0.1	&	94.8	&	0.4	&	0.01	&	0.1	&	94.8	&	0.4	\\
CC	&	1.21	&	1.21	&	0	&	0.2	&	1.21	&	1.21	&	0	&	0.2	\\
MLR	&	0.01	&	0.11	&	95	&	0.44	&	1.24	&	1.24	&	0	&	0.32	\\
PSPP	&	0.01	&	0.11	&	95.4	&	0.44	&	0.1	&	0.23	&	65.6	&	0.54	\\
PSBPP	&	0.01	&	0.11	&	95	&	0.44	&	0.32	&	0.36	&	65.4	&	1.31	\\
BART	&	0.25	&	0.27	&	29.8	&	0.36	&	0.25	&	0.27	&	29.8	&	0.36	\\
BARTps	&	0.23	&	0.25	&	43.4	&	0.41	&	0.23	&	0.25	&	43.4	&	0.41	\\
		\hline
		& \multicolumn{4}{c|}{Mean correct} & \multicolumn{4}{c|}{Both wrong}\\
		Method & Bias & RMSE & Coverage & AIL & Bias & RMSE & Coverage & AIL \\
		\hline
		BD	&	0.01	&	0.1	&	94.8	&	0.4	&	0.01	&	0.1	&	94.8	&	0.4	\\
CC	&	1.21	&	1.21	&	0	&	0.2	&	1.21	&	1.21	&	0	&	0.2	\\
MLR	&	0.01	&	0.11	&	95	&	0.44	&	1.24	&	1.24	&	0	&	0.32	\\
PSPP	&	0.01	&	0.11	&	95.8	&	0.44	&	0.53	&	0.58	&	41.6	&	0.91	\\
PSBPP	&	0.01	&	0.11	&	95	&	0.44	&	0.32	&	0.36	&	65.4	&	1.31	\\
BART	&	0.25	&	0.27	&	29.8	&	0.36	&	0.25	&	0.27	&	29.8	&	0.36	\\
BARTps	&	0.23	&	0.25	&	43.4	&	0.41	&	0.23	&	0.25	&	43.4	&	0.41	\\
		\hline
	\end{tabular}
\end{center}
\end{table}

\noindent
{\bf{Kang and Schafer (2007) example}}

\begin{table}
\caption{Bias, RMSE, 95\% coverage, and average 95\% confidence interval length (AIL) under the Kang and Schafer (2007) example with sample size 500 using bootstrap.\label{app_ks_500}}
\begin{center}
	%\tiny
	\hspace*{-1cm}
	\begin{tabular}{|lcccc|cccc|}
		\hline
		$n=500$ & \multicolumn{4}{c|}{Both correct} & \multicolumn{4}{c|}{Propensity correct} \\
		Method & Bias & RMSE & Coverage & AIL* & Bias & RMSE & Coverage & AIL \\ \hline
		BD	&	0.11	&	1.58	&	95.6	&	6.17	&	0.11	&	1.58	&	95.6	&	6.17	\\
CC	&	-9.96	&	10.2	&	0.2	&	8.42	&	-9.96	&	10.2	&	0.2	&	8.42	\\
MLR	&	0.03	&	1.55	&	99.2	&	9.04	&	-0.66	&	2.04	&	99.4	&	10.99	\\
PSPP	&	0.03	&	1.55	&	99.2	&	9.04	&	-0.11	&	1.71	&	99.6	&	9.66	\\
AIPWT	&	0.1	&	1.58	&	95.6	&	6.18	&	0.25	&	2.17	&	94	&	8.19	\\
PSBPP	&	0.03	&	1.55	&	99.2	&	9.04	&	1.72	&	2.54	&	97.4	&	11.18	\\
AIPWT with BART	&	-0.13	&	1.6	&	94.4	&	6.37	&	-0.6	&	1.75	&	91	&	6.74	\\
BART	&	-0.32	&	1.59	&	99.6	&	9.12	&	-1.05	&	1.94	&	98.6	&	9.36	\\
BARTps	&	-0.06	&	1.57	&	99.4	&	9.39	&	0.49	&	1.7	&	99	&	9.9	\\
		\hline
		& \multicolumn{4}{c|}{Mean correct} & \multicolumn{4}{c|}{Both wrong}\\
		Method & Bias & RMSE & Coverage & AIL & Bias & RMSE & Coverage & AIL \\	\hline
		BD	&	0.11	&	1.58	&	95.6	&	6.17	&	0.11	&	1.58	&	95.6	&	6.17	\\
CC	&	-9.96	&	10.2	&	0.2	&	8.42	&	-9.96	&	10.2	&	0.2	&	8.42	\\
MLR	&	0.03	&	1.55	&	99.2	&	9.04	&	-0.66	&	2.04	&	99.4	&	10.99	\\
PSPP	&	0.03	&	1.55	&	99.2	&	9.04	&	-1.99	&	2.77	&	92	&	9.42	\\
AIPWT	&	0.32	&	5.09	&	95.6	&	8	&	-46.4	&	858.18	&	68.6	&	326.68	\\
PSBPP	&	0.03	&	1.55	&	99.2	&	9.04	&	-1.39	&	2.43	&	98.8	&	11.89	\\
AIPWT with BART	&	-0.13	&	1.6	&	94.4	&	6.37	&	-0.75	&	1.81	&	90.8	&	6.74	\\
BART	&	-0.32	&	1.59	&	99.6	&	9.12	&	-1.05	&	1.94	&	98.6	&	9.36	\\
BARTps	&	-0.15	&	1.58	&	99.8	&	9.39	&	-0.89	&	1.87	&	99.2	&	9.75	\\
		\hline
	\end{tabular}
\end{center}
\end{table}

\begin{table}
\caption{Bias, RMSE, 95\% coverage, and average 95\% confidence interval length (AIL) under the Kang and Schafer (2007) example with sample size 1,000 using bootstrap.\label{app_ks_1000}}
\begin{center}
	%\tiny
	\hspace*{-1cm}
	\begin{tabular}{|lcccc|cccc|}
		\hline
		$n=1,000$ & \multicolumn{4}{c|}{Both correct} & \multicolumn{4}{c|}{Propensity correct} \\
		Method & Bias & RMSE & Coverage & AIL* & Bias & RMSE & Coverage & AIL \\ \hline
		BD	&	0.07	&	1.11	&	95.2	&	4.38	&	0.07	&	1.11	&	95.2	&	4.38	\\
CC	&	-9.96	&	10.09	&	0	&	5.97	&	-9.96	&	10.09	&	0	&	5.97	\\
MLR	&	0.07	&	1.11	&	99.4	&	6.38	&	-0.74	&	1.63	&	98	&	7.78	\\
PSPP	&	0.06	&	1.11	&	99.4	&	6.38	&	-0.07	&	1.21	&	99.2	&	6.66	\\
AIPWT	&	0.06	&	1.11	&	95.6	&	4.38	&	0.07	&	1.66	&	94.2	&	6.01	\\
PSBPP	&	0.07	&	1.11	&	99.4	&	6.38	&	1.46	&	1.95	&	96.8	&	7.4	\\
AIPWT with BART	&	-0.05	&	1.12	&	95.2	&	4.42	&	-0.31	&	1.19	&	93.8	&	4.61	\\
BART	&	-0.13	&	1.12	&	99.6	&	6.38	&	-0.59	&	1.29	&	99.2	&	6.5	\\
BARTps	&	0	&	1.11	&	99.4	&	6.46	&	0.39	&	1.23	&	99.2	&	6.8	\\
		\hline
		& \multicolumn{4}{c|}{Mean correct} & \multicolumn{4}{c|}{Both wrong}\\
		Method & Bias & RMSE & Coverage & AIL & Bias & RMSE & Coverage & AIL \\	\hline
		BD	&	0.07	&	1.11	&	95.2	&	4.38	&	0.07	&	1.11	&	95.2	&	4.38	\\
CC	&	-9.96	&	10.09	&	0	&	5.97	&	-9.96	&	10.09	&	0	&	5.97	\\
MLR	&	0.07	&	1.11	&	99.4	&	6.38	&	-0.74	&	1.63	&	98	&	7.78	\\
PSPP	&	0.07	&	1.11	&	99.4	&	6.38	&	-2.12	&	2.52	&	77.2	&	6.29	\\
AIPWT	&	-0.08	&	2.28	&	95.6	&	5.1	&	-35.69	&	477.13	&	41.2	&	196.51	\\
PSBPP	&	0.07	&	1.11	&	99.4	&	6.38	&	-1.13	&	1.73	&	99	&	7.84	\\
AIPWT with BART	&	-0.06	&	1.12	&	95.2	&	4.42	&	-0.45	&	1.24	&	93.2	&	4.62	\\
BART	&	-0.13	&	1.12	&	99.6	&	6.38	&	-0.59	&	1.29	&	99.2	&	6.5	\\
BARTps	&	-0.05	&	1.12	&	99.6	&	6.46	&	-0.52	&	1.27	&	99.2	&	6.7	\\
		\hline
	\end{tabular}
\end{center}
\end{table}

\begin{table}
\caption{Bias, RMSE, 95\% coverage, and average 95\% confidence interval length (AIL) under the Kang and Schafer (2007) example with sample size 5,000 using bootstrap.\label{app_ks_5000}}
\begin{center}
	%\tiny
	\hspace*{-1cm}
	\begin{tabular}{|lcccc|cccc|}
		\hline
		$n=5,000$ & \multicolumn{4}{c|}{Both correct} & \multicolumn{4}{c|}{Propensity correct} \\
		Method & Bias & RMSE & Coverage & AIL* & Bias & RMSE & Coverage & AIL \\ \hline
		BD	&	0.02	&	0.49	&	96.4	&	1.96	&	0.02	&	0.49	&	96.4	&	1.96	\\
CC	&	-9.94	&	9.97	&	0	&	2.65	&	-9.94	&	9.97	&	0	&	2.65	\\
MLR	&	0.02	&	0.5	&	99.4	&	2.87	&	-0.84	&	1.06	&	92	&	3.49	\\
PSPP	&	0.02	&	0.5	&	99.4	&	2.87	&	-0.04	&	0.53	&	99.2	&	2.95	\\
AIPWT	&	0.01	&	0.49	&	96.2	&	1.96	&	0.05	&	0.7	&	94.4	&	2.76	\\
PSBPP	&	0.02	&	0.5	&	99.4	&	2.87	&	0.86	&	1.01	&	88.8	&	3.07	\\
AIPWT with BART	&	-0.02	&	0.49	&	95.8	&	1.97	&	-0.08	&	0.51	&	95.6	&	2	\\
BART	&	-0.04	&	0.5	&	99.4	&	2.87	&	-0.24	&	0.56	&	99.2	&	2.89	\\
BARTps	&	-0.02	&	0.5	&	99.4	&	2.87	&	0.16	&	0.53	&	99.6	&	2.9	\\
		\hline
		& \multicolumn{4}{c|}{Mean correct} & \multicolumn{4}{c|}{Both wrong}\\
		Method & Bias & RMSE & Coverage & AIL & Bias & RMSE & Coverage & AIL \\	\hline
		BD	&	0.02	&	0.49	&	96.4	&	1.96	&	0.02	&	0.49	&	96.4	&	1.96	\\
CC	&	-9.94	&	9.97	&	0	&	2.65	&	-9.94	&	9.97	&	0	&	2.65	\\
MLR	&	0.02	&	0.5	&	99.4	&	2.87	&	-0.84	&	1.06	&	92	&	3.49	\\
PSPP	&	0.02	&	0.5	&	99.4	&	2.87	&	-2.28	&	2.36	&	5.2	&	2.62	\\
AIPWT	&	-0.01	&	0.59	&	96	&	2.07	&	-19.29	&	91.29	&	0.2	&	44.76	\\
PSBPP	&	0.02	&	0.5	&	99.4	&	2.87	&	-0.26	&	0.6	&	99.2	&	3.17	\\
AIPWT with BART	&	-0.02	&	0.49	&	95.8	&	1.97	&	-0.21	&	0.54	&	94.2	&	2	\\
BART	&	-0.04	&	0.5	&	99.4	&	2.87	&	-0.24	&	0.56	&	99.2	&	2.89	\\
BARTps	&	-0.02	&	0.5	&	99.4	&	2.87	&	-0.23	&	0.56	&	99.2	&	2.9	\\
		\hline
	\end{tabular}
\end{center}
\end{table}

\begin{table}
\caption{Bias, RMSE, 95\% coverage, and average 95\% confidence interval length (AIL) under the Kang and Schafer (2007) example with sample size 500 using MI with posterior mean of propensity scores.\label{app_ks_500_mi}}
\begin{center}
	%\tiny
	\hspace*{-1cm}
	\begin{tabular}{|lcccc|cccc|}
		\hline
		$n=500$ & \multicolumn{4}{c|}{Both correct} & \multicolumn{4}{c|}{Propensity correct} \\
		Method & Bias & RMSE & Coverage & AIL* & Bias & RMSE & Coverage & AIL \\ \hline
		BD	&	0.09	&	1.56	&	96.8	&	6.36	&	0.09	&	1.56	&	96.8	&	6.36	\\
CC	&	-10.02	&	10.25	&	0	&	6.11	&	-10.02	&	10.25	&	0	&	6.11	\\
MLR	&	0.08	&	1.56	&	96.4	&	6.34	&	-0.74	&	2.13	&	95	&	8.08	\\
PSPP	&	0.08	&	1.56	&	96.4	&	6.34	&	-0.06	&	1.74	&	95	&	6.69	\\
PSBPP	&	0.09	&	1.56	&	96.4	&	6.35	&	1.39	&	2.28	&	91.2	&	7.7	\\
BART	&	-0.15	&	1.58	&	96.4	&	6.36	&	-0.74	&	1.8	&	93.4	&	6.74	\\
BARTps	&	-0.05	&	1.58	&	97	&	6.47	&	0.35	&	1.7	&	97.4	&	6.97	\\
		\hline
		& \multicolumn{4}{c|}{Mean correct} & \multicolumn{4}{c|}{Both wrong}\\
		Method & Bias & RMSE & Coverage & AIL & Bias & RMSE & Coverage & AIL \\	\hline
		BD	&	0.09	&	1.56	&	96.8	&	6.36	&	0.09	&	1.56	&	96.8	&	6.36	\\
CC	&	-10.02	&	10.25	&	0	&	6.11	&	-10.02	&	10.25	&	0	&	6.11	\\
MLR	&	0.08	&	1.56	&	96.4	&	6.34	&	-0.74	&	2.13	&	95	&	8.08	\\
PSPP	&	0.08	&	1.56	&	96.4	&	6.34	&	-1.99	&	2.82	&	80.4	&	7.79	\\
PSBPP	&	0.08	&	1.56	&	96.4	&	6.35	&	-1.4	&	2.46	&	91.4	&	8.17	\\
BART	&	-0.15	&	1.58	&	96.4	&	6.36	&	-0.74	&	1.8	&	93.4	&	6.74	\\
BARTps	&	-0.08	&	1.58	&	96.8	&	6.48	&	-0.61	&	1.76	&	94.2	&	6.96	\\
		\hline
	\end{tabular}
\end{center}
\end{table}

\begin{table}
\caption{Bias, RMSE, 95\% coverage, and average 95\% confidence interval length (AIL) under the Kang and Schafer (2007) example with sample size 1,000 using MI with posterior mean of propensity scores.\label{app_ks_1000_mi}}
\begin{center}
	%\tiny
	\hspace*{-1cm}
	\begin{tabular}{|lcccc|cccc|}
		\hline
		$n=1,000$ & \multicolumn{4}{c|}{Both correct} & \multicolumn{4}{c|}{Propensity correct} \\
		Method & Bias & RMSE & Coverage & AIL* & Bias & RMSE & Coverage & AIL \\ \hline
		BD	&	0.05	&	1.11	&	96.2	&	4.49	&	0.05	&	1.11	&	96.2	&	4.49	\\
CC	&	-9.97	&	10.11	&	0	&	4.32	&	-9.97	&	10.11	&	0	&	4.32	\\
MLR	&	0.04	&	1.11	&	96.4	&	4.49	&	-0.82	&	1.68	&	90.8	&	5.69	\\
PSPP	&	0.05	&	1.11	&	96.4	&	4.49	&	-0.07	&	1.22	&	94.8	&	4.71	\\
PSBPP	&	0.05	&	1.11	&	96.6	&	4.49	&	0.99	&	1.61	&	91.2	&	5.27	\\
BART	&	-0.08	&	1.13	&	96.2	&	4.5	&	-0.46	&	1.24	&	95	&	4.71	\\
BARTps	&	-0.04	&	1.12	&	96.2	&	4.54	&	0.26	&	1.21	&	95.4	&	4.83	\\
		\hline
		& \multicolumn{4}{c|}{Mean correct} & \multicolumn{4}{c|}{Both wrong}\\
		Method & Bias & RMSE & Coverage & AIL & Bias & RMSE & Coverage & AIL \\	\hline
		BD	&	0.05	&	1.11	&	96.2	&	4.49	&	0.05	&	1.11	&	96.2	&	4.49	\\
CC	&	-9.97	&	10.11	&	0	&	4.32	&	-9.97	&	10.11	&	0	&	4.32	\\
MLR	&	0.04	&	1.11	&	96.4	&	4.49	&	-0.82	&	1.68	&	90.8	&	5.69	\\
PSPP	&	0.05	&	1.11	&	96.4	&	4.49	&	-2.17	&	2.57	&	65.4	&	5.34	\\
PSBPP	&	0.05	&	1.11	&	96.4	&	4.49	&	-1.41	&	1.92	&	83.8	&	5.55	\\
BART	&	-0.08	&	1.13	&	96.2	&	4.5	&	-0.46	&	1.24	&	95	&	4.71	\\
BARTps	&	-0.03	&	1.13	&	96.4	&	4.54	&	-0.4	&	1.24	&	95.4	&	4.85	\\
		\hline
	\end{tabular}
\end{center}
\end{table}

\begin{table}
\caption{Bias, RMSE, 95\% coverage, and average 95\% confidence interval length (AIL) under the Kang and Schafer (2007) example with sample size 5,000 using MI with posterior mean of propensity scores.\label{app_ks_5000_mi}}
\begin{center}
	%\tiny
	\hspace*{-1cm}
	\begin{tabular}{|lcccc|cccc|}
		\hline
		$n=5,000$ & \multicolumn{4}{c|}{Both correct} & \multicolumn{4}{c|}{Propensity correct} \\
		Method & Bias & RMSE & Coverage & AIL* & Bias & RMSE & Coverage & AIL \\ \hline
		BD	&	0.01	&	0.49	&	96.4	&	2.01	&	0.01	&	0.49	&	96.4	&	2.01	\\
CC	&	-9.94	&	9.96	&	0	&	1.93	&	-9.94	&	9.96	&	0	&	1.93	\\
MLR	&	0.01	&	0.49	&	96.4	&	2.01	&	-0.86	&	1.08	&	70.8	&	2.54	\\
PSPP	&	0.01	&	0.49	&	96.4	&	2.01	&	-0.03	&	0.53	&	97	&	2.1	\\
PSBPP	&	0.01	&	0.49	&	96	&	2.01	&	0.44	&	0.69	&	89	&	2.22	\\
BART	&	-0.03	&	0.49	&	96.6	&	2.01	&	-0.19	&	0.54	&	95.8	&	2.08	\\
BARTps	&	-0.02	&	0.5	&	96.8	&	2.01	&	0.1	&	0.51	&	96	&	2.08	\\
		\hline
		& \multicolumn{4}{c|}{Mean correct} & \multicolumn{4}{c|}{Both wrong}\\
		Method & Bias & RMSE & Coverage & AIL & Bias & RMSE & Coverage & AIL \\	\hline
		BD	&	0.01	&	0.49	&	96.4	&	2.01	&	0.01	&	0.49	&	96.4	&	2.01	\\
CC	&	-9.94	&	9.96	&	0	&	1.93	&	-9.94	&	9.96	&	0	&	1.93	\\
MLR	&	0.01	&	0.49	&	96.4	&	2.01	&	-0.86	&	1.08	&	70.8	&	2.54	\\
PSPP	&	0.01	&	0.49	&	96.4	&	2.01	&	-2.28	&	2.36	&	1.8	&	2.32	\\
PSBPP	&	0.01	&	0.49	&	96.4	&	2.01	&	-0.49	&	0.73	&	89.2	&	2.33	\\
BART	&	-0.03	&	0.49	&	96.6	&	2.01	&	-0.19	&	0.54	&	95.8	&	2.08	\\
BARTps	&	-0.02	&	0.5	&	97	&	2.01	&	-0.21	&	0.55	&	95.4	&	2.1	\\
		\hline
	\end{tabular}
\end{center}
\end{table}

\begin{table}
\caption{Bias, RMSE, 95\% coverage, and average 95\% confidence interval length (AIL) under the Kang and Schafer (2007) example with sample size 500 using MI with posterior draw of propensity scores.\label{app_ks_500_mi_new}}
\begin{center}
	%\tiny
	\hspace*{-1cm}
	\begin{tabular}{|lcccc|cccc|}
		\hline
		$n=500$ & \multicolumn{4}{c|}{Both correct} & \multicolumn{4}{c|}{Propensity correct} \\
		Method & Bias & RMSE & Coverage & AIL* & Bias & RMSE & Coverage & AIL \\ \hline
		BD	&	0.09	&	1.56	&	96.8	&	6.36	&	0.09	&	1.56	&	96.8	&	6.36	\\
CC	&	-10.02	&	10.25	&	0	&	6.11	&	-10.02	&	10.25	&	0	&	6.11	\\
MLR	&	0.08	&	1.56	&	96.4	&	6.34	&	-0.74	&	2.13	&	95	&	8.08	\\
PSPP	&	0.08	&	1.56	&	96.4	&	6.34	&	-0.11	&	1.72	&	96.6	&	7.22	\\
PSBPP	&	0.09	&	1.56	&	96.4	&	6.35	&	0.3	&	1.81	&	98	&	8.68	\\
BART	&	-0.15	&	1.58	&	96.4	&	6.36	&	-0.74	&	1.8	&	93.4	&	6.74	\\
BARTps	&	-0.12	&	1.58	&	97	&	6.5	&	-0.39	&	1.67	&	96.4	&	7.08	\\
		\hline
		& \multicolumn{4}{c|}{Mean correct} & \multicolumn{4}{c|}{Both wrong}\\
		Method & Bias & RMSE & Coverage & AIL & Bias & RMSE & Coverage & AIL \\	\hline
		BD	&	0.09	&	1.56	&	96.8	&	6.36	&	0.09	&	1.56	&	96.8	&	6.36	\\
CC	&	-10.02	&	10.25	&	0	&	6.11	&	-10.02	&	10.25	&	0	&	6.11	\\
MLR	&	0.08	&	1.56	&	96.4	&	6.34	&	-0.74	&	2.13	&	95	&	8.08	\\
PSPP	&	0.09	&	1.56	&	96.6	&	6.35	&	-1.94	&	2.73	&	84.2	&	8.24	\\
PSBPP	&	0.08	&	1.56	&	96.4	&	6.35	&	-1.22	&	2.26	&	96	&	8.88	\\
BART	&	-0.15	&	1.58	&	96.4	&	6.36	&	-0.74	&	1.8	&	93.4	&	6.74	\\
BARTps	&	-0.13	&	1.58	&	97	&	6.47	&	-0.76	&	1.81	&	93.8	&	6.95	\\
		\hline
	\end{tabular}
\end{center}
\end{table}

\begin{table}
\caption{Bias, RMSE, 95\% coverage, and average 95\% confidence interval length (AIL) under the Kang and Schafer (2007) example with sample size 1,000 using MI with posterior draw of propensity scores.\label{app_ks_1000_mi_new}}
\begin{center}
	%\tiny
	\hspace*{-1cm}
	\begin{tabular}{|lcccc|cccc|}
		\hline
		$n=1,000$ & \multicolumn{4}{c|}{Both correct} & \multicolumn{4}{c|}{Propensity correct} \\
		Method & Bias & RMSE & Coverage & AIL* & Bias & RMSE & Coverage & AIL \\ \hline
		BD	&	0.05	&	1.11	&	96.2	&	4.49	&	0.05	&	1.11	&	96.2	&	4.49	\\
CC	&	-9.97	&	10.11	&	0	&	4.32	&	-9.97	&	10.11	&	0	&	4.32	\\
MLR	&	0.04	&	1.11	&	96.4	&	4.49	&	-0.82	&	1.68	&	90.8	&	5.69	\\
PSPP	&	0.05	&	1.11	&	96.4	&	4.49	&	-0.12	&	1.21	&	96.4	&	5.03	\\
PSBPP	&	0.05	&	1.11	&	96.4	&	4.49	&	0.07	&	1.26	&	98.2	&	6.14	\\
BART	&	-0.08	&	1.13	&	96.2	&	4.5	&	-0.46	&	1.24	&	95	&	4.71	\\
BARTps	&	-0.07	&	1.12	&	96.4	&	4.54	&	-0.22	&	1.18	&	96.6	&	4.91	\\
		\hline
		& \multicolumn{4}{c|}{Mean correct} & \multicolumn{4}{c|}{Both wrong}\\
		Method & Bias & RMSE & Coverage & AIL & Bias & RMSE & Coverage & AIL \\	\hline
		BD	&	0.05	&	1.11	&	96.2	&	4.49	&	0.05	&	1.11	&	96.2	&	4.49	\\
CC	&	-9.97	&	10.11	&	0	&	4.32	&	-9.97	&	10.11	&	0	&	4.32	\\
MLR	&	0.04	&	1.11	&	96.4	&	4.49	&	-0.82	&	1.68	&	90.8	&	5.69	\\
PSPP	&	0.05	&	1.11	&	96.4	&	4.49	&	-2.12	&	2.52	&	68.6	&	5.47	\\
PSBPP	&	0.05	&	1.11	&	96.6	&	4.49	&	-1.42	&	1.92	&	89.4	&	6.25	\\
BART	&	-0.08	&	1.13	&	96.2	&	4.5	&	-0.46	&	1.24	&	95	&	4.71	\\
BARTps	&	-0.07	&	1.12	&	96.4	&	4.54	&	-0.47	&	1.25	&	96.2	&	4.85	\\
		\hline
	\end{tabular}
\end{center}
\end{table}

\begin{table}
\caption{Bias, RMSE, 95\% coverage, and average 95\% confidence interval length (AIL) under the Kang and Schafer (2007) example with sample size 5,000 using MI with posterior draw of propensity scores.\label{app_ks_5000_mi_new}}
\begin{center}
	%\tiny
	\hspace*{-1cm}
	\begin{tabular}{|lcccc|cccc|}
		\hline
		$n=5,000$ & \multicolumn{4}{c|}{Both correct} & \multicolumn{4}{c|}{Propensity correct} \\
		Method & Bias & RMSE & Coverage & AIL* & Bias & RMSE & Coverage & AIL \\ \hline
		BD	&	0.01	&	0.49	&	96.4	&	2.01	&	0.01	&	0.49	&	96.4	&	2.01	\\
CC	&	-9.94	&	9.96	&	0	&	1.93	&	-9.94	&	9.96	&	0	&	1.93	\\
MLR	&	0.01	&	0.49	&	96.4	&	2.01	&	-0.86	&	1.08	&	70.8	&	2.54	\\
PSPP	&	0.01	&	0.49	&	96.4	&	2.01	&	-0.04	&	0.53	&	97.6	&	2.21	\\
PSBPP	&	0.01	&	0.49	&	96.2	&	2.01	&	-0.17	&	0.55	&	98.6	&	2.53	\\
BART	&	-0.03	&	0.49	&	96.6	&	2.01	&	-0.19	&	0.54	&	95.8	&	2.08	\\
BARTps	&	-0.02	&	0.5	&	96.8	&	2.02	&	-0.08	&	0.51	&	96.8	&	2.1	\\
		\hline
		& \multicolumn{4}{c|}{Mean correct} & \multicolumn{4}{c|}{Both wrong}\\
		Method & Bias & RMSE & Coverage & AIL & Bias & RMSE & Coverage & AIL \\	\hline
		BD	&	0.01	&	0.49	&	96.4	&	2.01	&	0.01	&	0.49	&	96.4	&	2.01	\\
CC	&	-9.94	&	9.96	&	0	&	1.93	&	-9.94	&	9.96	&	0	&	1.93	\\
MLR	&	0.01	&	0.49	&	96.4	&	2.01	&	-0.86	&	1.08	&	70.8	&	2.54	\\
PSPP	&	0.01	&	0.49	&	96.4	&	2.01	&	-2.27	&	2.34	&	2.8	&	2.35	\\
PSBPP	&	0.01	&	0.49	&	96	&	2.01	&	-1	&	1.13	&	74.6	&	2.7	\\
BART	&	-0.03	&	0.49	&	96.6	&	2.01	&	-0.19	&	0.54	&	95.8	&	2.08	\\
BARTps	&	-0.02	&	0.5	&	96.8	&	2.02	&	-0.2	&	0.54	&	95.8	&	2.11	\\
		\hline
	\end{tabular}
\end{center}
\end{table}

\noindent
{\bf{Web Appendix D: Simple descriptive statistics for NASS-CDS 2014 and FARS 2015}}

\noindent
{\bf{NASS-CDS 2014}}

\begin{table}
	\caption{Summary statistics stratified by missingness in total delta-v.\label{des_cds1}}
	\centering
	\begin{tabular}{|l|r|r|r|}
		\hline
		Variables & Missing (\%) & Non-missing (\%) & $p$-value \\
		\hline
		Crash type	&		&		&	$<0.01$	\\
		$\quad$ Change traffic-way, vehicle turning	&	18.4	&	32.3	&		\\
		$\quad$ Same traffic-way	&	29.3	&	32.8	&		\\
		$\quad$ Single driver	&	31.1	&	11.5	&		\\
		$\quad$ Others or missing	&	21.3	&	23.4	&		\\
		Heading angle	&		&		&	$<0.01$	\\
		$\quad$ Frontal	&	17.8	&	24.6	&		\\
\hline
	\end{tabular}
\end{table}

\begin{table}
	\caption{Summary statistics stratified by missingness in total delta-v, continued.\label{des_cds2}}
	\centering
	\begin{tabular}{|l|r|r|r|}
		\hline
		Variables & Missing (\%) & Non-missing (\%) & $p$-value \\
		\hline
		Vehicle make &		&		&	0.11	\\
		$\quad$ American	&	47.2	&	50.6	&		\\
		$\quad$ Japanese	&	39.6	&	36.3	&		\\
		$\quad$ Europe or other foreign	&	13.2	&	13.1	&		\\
		Avoidance maneuver?	&		&		&	$<0.01$	\\
		$\quad$ Yes	&	18.8	&	23.6	&		\\
		$\quad$ No	&	35.8	&	36.7	&		\\
		$\quad$ Missing	&	45.4	&	39.8	&		\\
		Model year	&		&		&	0.16	\\
		$\quad$ $<2003$ or Missing	&	33.7	&	31.5	&		\\
		$\quad$ $\geq 2003$	&	66.3	&	68.5	&		\\
		No. of occupants &		&		&	0.81	\\
		$\quad$ 1	&	71.3	&	70.6	&		\\
		$\quad$ 2	&	19.1	&	19.9	&		\\
		$\quad$ $\geq 3$	&	9.6	&	9.5	&		\\
		Pre-crash event &		&		&	$<0.01$	\\
		$\quad$ Traveling	&	42.1	&	35.1	&		\\
		$\quad$ Loss control	&	9.3	&	5.7	&		\\
		$\quad$ Other or Missing	&	48.6	&	59.3	&		\\
		Pre-event movement	&		&		&	0.59	\\
		$\quad$ Going straight	&	55.6	&	54.6	&		\\
		$\quad$ Other or Missing	&	44.4	&	45.4	&		\\
		Pre-impact stability	&		&		&	$<0.01$	\\
		$\quad$ Skidding	&	10.6	&	9.5	&		\\
		$\quad$ Tracking	&	74.2	&	79	&		\\
		$\quad$ Other or Missing	&	15.3	&	11.5	&		\\
		Road alignment	&		&		&	0.31	\\
		$\quad$ Straight	&	79.5	&	80.7	&		\\
		$\quad$ Curve left	&	10.6	&	9.1	&		\\
		$\quad$ Curve right	&	9.9	&	10.2	&		\\
		Surface condition	&		&		&	0.03	\\
		$\quad$ Dry	&	81.4	&	79.6	&		\\
		$\quad$ Wet	&	12.6	&	15.5	&		\\
		$\quad$ Other or Missing	&	5.9	&	4.9	&		\\
		Surface type	&		&		&	0.07	\\
		$\quad$ Concrete	&	13.4	&	11.4	&		\\
		$\quad$ Asphalt and Others	&	86.6	&	88.6	&		\\
		Race	&		&		&	$<0.01$	\\
		$\quad$ White	&	34.8	&	38.4	&		\\
		$\quad$ Black	&	10.3	&	12.3	&		\\
		$\quad$ Other or Missing	&	54.9	&	49.3	&		\\
		Relation to interchange	&		&		&	$<0.01$	\\
		$\quad$ Interchange area related	&	12.7	&	11.5	&		\\
		$\quad$ Intersection related	&	41.4	&	57.5	&		\\
		$\quad$ Non-interchange area and non-junction	&	45.9	&	31	&		\\
		\hline
	\end{tabular}
\end{table}

\begin{table}
	\caption{Summary statistics stratified by missingness in total delta-v, continued.\label{des_cds3}}
	\centering
	\begin{tabular}{|l|r|r|r|}
		\hline
		Variables & Missing (\%) & Non-missing (\%) & $p$-value \\
		\hline
		Other drug test results	&		&		&	$<0.01$	\\
		$\quad$ No test given	&	80.9	&	87.2	&		\\
		$\quad$ Drugs found	&	1.7	&	2.6	&		\\
		$\quad$ Drugs not found	&	3	&	2	&		\\
		$\quad$ Results not known	&	3.2	&	1.9	&		\\
		$\quad$ Missing	&	11.1	&	6.2	&		\\
		Traffic control device	&		&		&	$<0.01$	\\
\hline
	\end{tabular}
\end{table}

\begin{table}
	\caption{Summary statistics stratified by missingness in total delta-v, continued.\label{des_cds4}}
	\centering
	\begin{tabular}{|l|r|r|r|}
		\hline
		Variables & Missing (\%) & Non-missing (\%) & $p$-value \\
		\hline
		Highest deformation extent	&		&		&	$<0.01$	\\
		$\quad$ 1	&	12.8	&	26.3	&		\\
		$\quad$ $\geq 2$	&	32.5	&	60.9	&		\\
		$\quad$ Missing	&	54.7	&	12.8	&		\\
		Second highest deformation extent	&		&		&	$<0.01$	\\
		$\quad$ 1	&	8.8	&	15	&		\\
		$\quad$ $\geq 2$	&	8.8	&	10.6	&		\\
\hline
	\end{tabular}
\end{table}

\begin{table}
	\caption{Summary statistics stratified by missingness in total delta-v, continued.\label{des_cds5}}
	\centering
	\begin{tabular}{|l|r|r|r|}
		\hline
		Variables & Missing (\%) & Non-missing (\%) & $p$-value \\
		\hline
		Police reported belt use	&		&		&	0.01	\\
		$\quad$ None used	&	8.1	&	6.6	&		\\
		$\quad$ Used	&	82.1	&	85.7	&		\\
		$\quad$ Not reported or Missing	&	9.8	&	7.6	&		\\
		Sex	&		&		&	$<0.01$	\\
		$\quad$ Female	&	40.4	&	46	&		\\
		$\quad$ Male or Missing	&	59.6	&	54	&		\\
		Driver's weight &		&		&	0.84	\\
		$\quad$ $<60$kg	&	6.4	&	6.9	&		\\
		$\quad$ 60-70kg	&	9	&	9.6	&		\\
		$\quad$ 70-80kg	&	11.7	&	10.5	&		\\
		$\quad$ 80-90kg	&	8.2	&	9	&		\\
		$\quad$ 90-100kg	&	6	&	6.2	&		\\
		$\quad$ $\geq 100$kg	&	8.4	&	8.2	&		\\
		$\quad$ Missing	&	50.3	&	49.6	&		\\
		\hline
	\end{tabular}
\end{table}

\noindent
{\bf{2015 FARS}}

\begin{table}
	\caption{Summary statistics stratified by missingness in blood alcohol concentration (BAC)\label{des_fars1}}
	\small
	\centering
	\begin{tabular}{|l|r|r|r|}
		\hline
		Variables & Missing (\%) & Non-missing (\%) & $p$-value \\
		\hline
		Hour of crash	&		&		&	$<0.01$	\\
		$\quad$ 12-6am	&	15.5	&	23.1	&		\\
		$\quad$ 6-10am	&	13.1	&	13.4	&		\\
		$\quad$ 10am-4pm	&	26.8	&	23.5	&		\\
		$\quad$ 4-8pm	&	24.3	&	19.9	&		\\
		$\quad$ 8pm-12am	&	20	&	19.4	&		\\
		$\quad$ Unknown	&	0.4	&	0.7	&		\\
\hline
	\end{tabular}
\end{table}

\begin{table}
	\caption{Summary statistics stratified by missingness in blood alcohol concentration (BAC), continued \label{des_fars2}}
	\small
	\centering
	\begin{tabular}{|l|r|r|r|}
		\hline
		Variables & Missing (\%) & Non-missing (\%) & $p$-value \\
		\hline
		Age	&		&		&	$<0.01$	\\
		$\quad$ Younger than 21	&	12.3	&	12.8	&		\\
		$\quad$ 21-30	&	24.6	&	28.7	&		\\
		$\quad$ 30-40	&	16.1	&	17	&		\\
		$\quad$ 40-50	&	11.8	&	10.7	&		\\
		$\quad$ 50-60	&	12.2	&	11.7	&		\\
		$\quad$ Older than 60	&	21.4	&	19.1	&		\\
\hline
	\end{tabular}
\end{table}

\begin{table}
	\caption{Summary statistics stratified by missingness in blood alcohol concentration (BAC), continued \label{des_fars3}}
	\small
	\centering
	\begin{tabular}{|l|r|r|r|}
		\hline
		Variables & Missing (\%) & Non-missing (\%) & $p$-value \\
		\hline
		Underride?	&		&		&	$<0.01$	\\
		$\quad$ Yes or unknown	&	0.7	&	1.7	&		\\
		$\quad$ No 	&	99.3	&	98.3	&		\\
		Rollover location	&		&		&	$<0.01$	\\
		$\quad$ No rollover	&	91.7	&	80.1	&		\\
		$\quad$ On roadside	&	5.6	&	14.2	&		\\
		$\quad$ Other or unknown	&	2.6	&	5.8	&		\\
\hline
	\end{tabular}
\end{table}

\begin{table}
	\caption{Summary statistics stratified by missingness in blood alcohol concentration (BAC), continued \label{des_fars4}}
	\footnotesize
	\centering
	\begin{tabular}{|l|r|r|r|}
		\hline
		Variables & Missing (\%) & Non-missing (\%) & $p$-value \\
		\hline
		Speed limit	&		&		&	$<0.01$	\\
		$\quad$ $\leq 25$ mph	&	5	&	4.5	&		\\
		$\quad$ 30 mph	&	5.4	&	4.6	&		\\
		$\quad$ 35 mph	&	11.4	&	10	&		\\
		$\quad$ 40 mph	&	8.7	&	7.3	&		\\
		$\quad$ 45 mph	&	17.2	&	15	&		\\
		$\quad$ 50 mph	&	4.9	&	5.4	&		\\
\hline
	\end{tabular}
\end{table}

\label{lastpage}

\end{document}